\documentclass[twocolumn,amsmath,amssymb,prd,10pt,floatfix, 
superscriptaddress,nofootinbib, showpacs, preprintnumbers]{revtex4-1}

\usepackage{graphicx}
\usepackage{dcolumn}
\usepackage{bbold}
\usepackage{amsmath}
\usepackage{amsfonts}
\usepackage{color}
\usepackage[version=0.96]{pgf}
\usepackage[latin1]{inputenc}
\usepackage{verbatim}
\usepackage{footnote}
\usepackage{slashed}

\newcommand{\estate}[1]{\ensuremath{\mathfrak{#1}}}

\begin{document}

\preprint{JLAB-THY-15-2004}

\title{ Excited meson radiative transitions from lattice QCD \\using variationally optimized operators}

 \author{Christian~J.~Shultz}
 \email{shultz@jlab.org}
 \affiliation{Department of Physics, Old Dominion University, Norfolk, VA 23529, USA}

 \author{Jozef~J.~Dudek}
 \email{dudek@jlab.org}
 \affiliation{Department of Physics, Old Dominion University, Norfolk, VA 23529, USA}
 \affiliation{Theory Center, Jefferson Lab, 12000 Jefferson Avenue, Newport News, VA 23606, USA}

 \author{Robert~G.~Edwards}
 \email{edwards@jlab.org}
 \affiliation{Theory Center, Jefferson Lab, 12000 Jefferson Avenue, Newport News, VA 23606, USA}

\collaboration{for the Hadron Spectrum Collaboration}
\noaffiliation


\date{\today}

\begin{abstract}
We explore the use of `optimized' operators, designed to interpolate only a single meson eigenstate, in three-point correlation functions with a vector-current insertion. These operators are constructed as linear combinations in a large basis of meson interpolating fields using a variational analysis of matrices of two-point correlation functions. After performing such a determination at both zero and non-zero momentum, we compute three-point functions and are able to study radiative transition matrix elements featuring excited state mesons. The required two- and three-point correlation functions are efficiently computed using the \emph{distillation} framework in which there is a factorization between quark propagation and operator construction, allowing for a large number of meson operators of definite momentum to be considered. We illustrate the method with a calculation using anisotopic lattices having three flavors of dynamical quark all tuned to the physical strange quark mass, considering form-factors and transitions of pseudoscalar and vector meson excitations. The dependence on photon virtuality for a number of form-factors and transitions is extracted and some discussion of excited-state phenomenology is presented.
\end{abstract}

\maketitle

\section{Introduction\label{sec::intro}}

The coupling of mesons and baryons to photons at leading order in $\alpha_{\mathrm{em}}$ is given by matrix elements of the quark-field vector current ${\big\langle f \big| j^\mu \big| i  \big\rangle}$ where ${j^\mu(x) = \bar{\psi}(x) \gamma^\mu \psi(x)}$. In the case that the initial and final state hadron are the same we express the matrix element in terms of Lorentz-invariant form-factors, whose dependence on the virtuality of the photon, $Q^2$, can be related to quark charge and current distributions within the hadron. The current can also induce a transition from one hadron eigenstate $\big| i \big\rangle$ to another $\big|f \big\rangle$, in which case we speak of transition form-factors. For hadrons with non-zero spin, there are multiple possible amplitudes which can be labeled by the helicity of the hadrons, or we may expand the current in terms of multipoles to provide another convenient physically motivated basis.

In the meson sector, these matrix elements appear in photo-- and electro--production of mesons from nucleon and nuclear targets, where the coupling of the photon to a $t$-channel meson exchange is described by transition form-factors. Measurements of these processes with unprecedented statistics will be made in the GlueX and CLAS12 detectors at the 12 GeV upgraded CEBAF \cite{Dudek:2012vr}. In particular, photoproduction has been proposed as a means to produce large numbers of exotic $J^{PC}$ hybrids, those mesons which contain an excitation of the gluonic field as well as the usual quark-antiquark pair \cite{Horn:1977rq, *Barnes:1982tx, *Chanowitz:1982qj, *Isgur:1985vy}, \cite{Dudek:2011bn}. 

Production of excited mesons at very forward angles using pion beams is also driven by the vector-current transition form-factor. In the Primakoff process the pion absorbs a nearly on-shell photon from a nucleon or nuclear target, with an amplitude to transition to another meson species given by transition form-factors at $Q^2 \approx 0$. Recent such measurements of the couplings $a_2 \to \pi \gamma$ and $\pi_2 \to \pi \gamma$ have been made at COMPASS \cite{Adolph:2014mup}. 

In charmonium and bottomonium, the relatively small total widths of the low-lying states means that radiative transition rates between them constitute significant branching fractions, and can be measured directly, as can rates of decay to a photon plus light-quark mesons where the heavy quark-antiquark pair annihilates \cite{PDG-2012, Ablikim:2011kv, *Ablikim:2011da, *Ablikim:2012sf, *Ablikim:2013hq}.

In the baryon sector, the $Q^2$ dependence of transition form-factors can be measured quite directly in electro-production of excited nucleons off proton and neutron targets. The relative magnitudes of the various multipole or helicity amplitudes and the variation with photon virtuality have been discussed as a means to study the internal quark-gluon structure of excited $N^\star$ and $\Delta^\star$ states \cite{Aznauryan:2012ba}. 

The properties of hadrons constructed from strongly interacting quarks and gluons should be calculable within the relevant gauge field theory, Quantum Chromodynamics (QCD). At the energy scale of hadrons, QCD does not have a small coupling constant and must be treated non-perturbatively. The tool we will use to achieve this is lattice QCD, in which the field theory is discretised on a finite grid of Euclidean space-time points, and where we can compute correlation functions as an average over a finite but large number of possible gauge-field configurations.

The vector current matrix elements that we are interested in, $\big\langle {f} \big| j^\mu \big| {i} \big\rangle$, appear in three-point correlation functions of generic form, $\big\langle 0 \big| \mathcal{O}_{f}(t_f) \, j^\mu(t)\,  \mathcal{O}_{i}^\dag(t_i) \big| 0 \big\rangle$. Here $\mathcal{O}_{i}^\dag$, $\mathcal{O}_{f}^\dag$ are operators constructed from quark and gluon fields, capable of interpolating mesons from the vacuum. In general, such hadron operators with definite quantum numbers interpolate not just one QCD eigenstate, but rather a linear superposition of all states with those quantum numbers. Each state propagates through Euclidean time with a factor $e^{-E t}$ such that at large times only the state of lowest energy survives, suggesting that if we separate the three operators in the correlation function by large time intervals we will obtain the transition between the lightest states with the quantum numbers of $\mathcal{O}_{i}$, $\mathcal{O}_{f}$. However in many cases we actually wish to study states which are not the lightest with a given set of quantum numbers. If we use a generic operator to interpolate them from the vacuum their contribution will be as subleading exponential dependences in the correlation function, which become dominated by lighter states as we propagate in Euclidean time. Determining the amplitude of subleading exponential contributions through fitting the time-dependence proves to be unreliable.

Our solution to this problem is to form `optimal' operators for interpolation of each state that we wish to study, that is operators which have a dominant amplitude to produce a particular state, and significantly reduced amplitude to produce all other states, particularly those lighter than the state in question. In this case the three-point correlation functions are dominated by the desired initial and final states, even if they are not the lightest in a given quantum number sector. Optimal operators can be constructed as a linear superposition of operators in a large basis, where there is a particular linear superposition for each state in the tower of excited states. The different superpositions are orthogonal in a suitable sense. We are able to find appropriate superpositions through a variational analysis of the matrix of \emph{two-point} correlation functions, $\big\langle 0 \big| \mathcal{O}^{}_i(t) \mathcal{O}^\dag_j(0) \big| 0 \big\rangle$, for a set of operators $\{ \mathcal{O}_i \}$.

The technology we will utilize, constructing `optimized' operators as linear superpositions of a large basis of interpolators, is useful not only for extracting transitions featuring excited states, but also to improve ground-state signals by significantly reducing the unwanted contributions of excited states to correlation functions. The presence of such contributions remains a problem for calculations attempting precision extraction of ground-state matrix elements \cite{Aznauryan:2012ba,PhysRevD.83.045010,Horsley:2013ayv,PhysRevLett.96.052001,Lin:2012ev,Bali:2014gha,Dinter:2011sg,Green:2014xba}. Increase in statistical noise precludes separating operators by large time separations, so instead the use of `optimized' operators seeks to deal with the problem directly by suppressing creation of the unwanted excited states. 

The use of optimized operators as a tool to extract excited-state radiative transitions was previously considered in \cite{PhysRevD.79.094504}, where the meson source operators, $\mathcal{O}^\dag_{i}$ were optimized, while the meson sink operators, $\mathcal{O}_{f}$, were simple local fermion bilinears. Quark propagation from the meson sink proceeded using the `sequential source' method, which proved to be a significant limitation on what could be achieved. In this study we will make use of the \emph{distillation} framework for correlator construction \cite{Peardon:2009gh} within which we will be able to use optimized operators of definite momentum at both source and sink as well as to insert a vector current operator of definite momentum. This is the first use of distillation to compute three-point functions.

In order to demonstrate the technology we perform a calculation on dynamical lattices having three flavors of quark all tuned to approximately the physical strange quark mass - the spectrum of isovector mesons on these lattices was previously presented in \cite{Dudek:2009qf, Dudek:2010wm}. We consider pseudoscalar and vector mesons in the SU(3) flavor octet, extracting form-factors and transition form-factors for both ground and excited states in these channels. This should be expected to be a challenging undertaking -- transitions between excited-state and ground-state vector and pseudoscalar mesons are `hindered' and have relatively small magnitude for small photon virtualities -- as such we will be extracting small signals from correlation functions built using `optimal' excited state operators.

In section \ref{sec::formfactor} we introduce the decomposition of vector current matrix elements in terms of Lorentz covariant kinematic factors multiplying the unknown form-factors we seek to extract. We proceed to introduce the gauge configurations, operator basis, and background information relevant to two-point calculations in Sections \ref{sec::calc_details} and \ref{sec::two_points}.  Section \ref{sec::three-point} concerns three-point functions; we describe their calculation within the distillation framework and their relation to the matrix elements of interest while also demonstrating the efficacy of optimized operators. We present the results of our calculation in Section \ref{sec::results}, comparing with relevant previous calculations, and then conclude with a summary and outlook in Section \ref{sec::summary}. Appendices describing helicity operators, momentum conservation in a finite-volume, the improvement of the vector current and the flavor structure of the current follow.

\section{Form-factors and Transitions \label{sec::formfactor}}

The photon couples to the electric charges of $u,d,s$ quarks via the vector current ${j^\mu = +\tfrac{2}{3}\, \bar{u} \gamma^\mu u - \tfrac{1}{3}\, \bar{d} \gamma^\mu d - \tfrac{1}{3}\, \bar{s} \gamma^\mu s }$, up to a factor of the magnitude of the electron charge, $e$. In general a transition induced by this current between a hadron, $h$, of spin-$J$ and a hadron $h'$ of spin-$J'$ is described by the matrix element,
\begin{equation}
\big\langle h'_{J'}(\lambda',\vec{p}\,') \big| j^\mu \big| h_J(\lambda,\vec{p}\,)\big\rangle, \nonumber
\end{equation}
where the spin-state of $h$ is specified in terms of its helicity, $\lambda$, the projection of $\vec{J}$ along the direction of momentum $\vec{p}$. These matrix elements are simply related to the helicity amplitude for the transition $\gamma\, h \to h'$ by including the initial-state photon's polarization vector,
\begin{multline}
\mathcal{M}\big( \gamma(\lambda_\gamma,\vec{q}\,)\, h_J(\lambda,\vec{p}\, ) \to h'_{J'}(\lambda,\vec{p}\,')  \big) \\
= \epsilon_\mu(\lambda_\gamma, \vec{q}\,) \big\langle h'_{J'}(\lambda',\vec{p}\,') \big| j^\mu \big| h_J(\lambda,\vec{p}\,)\big\rangle, \nonumber
\end{multline}
where $\vec{q} = \vec{p}\,' - \vec{p}$ and where the photon has a virtuality ${Q^2 = -q^2 = \big| \vec{p}\,' - \vec{p} \,\big|^2 - \big(E_{h'}(\vec{p}\,') - E_h(\vec{p}\,)\, \big)^2 }$.

There are relations between these amplitudes which follow from the constraints of Lorentz invariance, current conservation and invariance under parity transformations. These can be accounted for if we write a matrix-element decomposition in terms of a number of Lorentz invariant form-factors, $F_i(Q^2)$,
\begin{multline}
\big\langle h'_{J'}(\lambda',\vec{p}\,') \big| j^\mu \big| h_J(\lambda,\vec{p})\big\rangle \\
= \sum\nolimits_i \, K_i^\mu \big[   h'_{J'}\big(\lambda',\vec{p}\,'); h_{J}(\lambda,\vec{p}) \big] \, F_i(Q^2). \label{decomp}
\end{multline}
The Lorentz covariant `kinematic factors', $K_i^\mu$, are constructed from the meson four-momenta, $p_\nu$, $p_\nu'$, and initial and final state polarization tensors relevant to the spin of the mesons, $\epsilon_{\rho \sigma \ldots}(\lambda, \vec{p})$, $\epsilon^*_{\rho \sigma \ldots}(\lambda', \vec{p}\,')$. For any given pair of mesons $h,h'$, of definite spin and parity, there are only a limited number of possible constructions consistent with parity invariance and with the additional constraint of current conservation, we can write explicit decompositions in terms of a few independent form-factors. 

For example, a pseudoscalar particle like the pion has only a single form-factor appearing in its decomposition,
\begin{equation}
\big\langle \pi^+(\vec{p}\,') \big| j^\mu \big| \pi^+(\vec{p}) \big\rangle = (p+p')^\mu \, F_\pi(Q^2).
\label{pion_ff}
\end{equation} 
At $Q^2=0$, the vector current measures the charge of the pion in units of $e$, so $F_\pi(0) = 1$ exactly. 

In a transition between two different pseudoscalar particles, there is again only one form-factor, but the kinematic factor differs owing to the differing masses ($m, m'$) of the pseudoscalar particles ($\pi, \pi'$),
\begin{align}
\big\langle \pi'^+ & (\vec{p}\,') \big| j^\mu \big| \pi^+(\vec{p}) \big\rangle \nonumber \\
&= \left[ (p\!+\!p')^\mu \!+\! \tfrac{m'^2 - m^2}{Q^2} (p' \!-\! p)^\mu \right]\!  F_{\pi'\pi}(Q^2). \label{pi_pi_ff}
\end{align} 

The transition matrix-element between a vector particle and a pseudoscalar can be expressed as
\begin{align}
&\big\langle \pi^+(\vec{p}\,') \big| j^\mu \big| \rho^+(\lambda, \vec{p}) \big\rangle  \notag \\
&\qquad= \epsilon^{\mu \nu \rho \sigma} p'_\nu\, p^{\,}_\rho \,\epsilon^{\,}_\sigma(\lambda, \vec{p}\,) \, \tfrac{2}{m_\pi + m_\rho} F_{\rho \pi}(Q^2),
\label{rho_pi_ff}
\end{align} 
and for a vector meson stable under the strong interactions, the transition form-factor at ${Q^2=0}$ can be related to the radiative decay width ${\Gamma( \rho^+ \to \pi^+ \gamma) = \frac{4}{3} \alpha \frac{\lvert \vec{q}\,  \rvert^3}{(m_\rho + m_\pi)^2} \lvert F_{\rho \pi}(0) \rvert^2    }$, where $\vec{q}$ is the momentum of the final-state photon in the rest-frame of the decaying $\rho$ meson.

The vector current matrix element for a stable vector hadron of mass $m$ has three possible covariant structures having the right parity transformation properties once current conservation is demanded \cite{Arnold:1979cg}, 
\begin{align}
\big\langle \rho&^+(\lambda',\vec{p}\,') \big| j^\mu \big| \rho^+(\lambda, \vec{p}) \big\rangle \nonumber \\
= 	&- \big[(p+p')^\mu \; \epsilon^*(\lambda',\vec{p}\,') \cdot \epsilon(\lambda,\vec{p}) \big]\; G_1(Q^2) \nonumber \\
	&+ \big[\epsilon^\mu(\lambda,\vec{p}) \, \epsilon^*(\lambda',\vec{p}\,') \!\cdot\! p + \epsilon^{\mu*}(\lambda',\vec{p}\,')\, \epsilon(\lambda,\vec{p})\!\cdot\! p' \big]\, G_2(Q^2) \nonumber \\
	&- \big[(p+p')^\mu \; \epsilon^*(\lambda',\vec{p}\,') \!\cdot\! p \;  \epsilon(\lambda,\vec{p})\!\cdot\! p' \, \tfrac{1}{2m^2} \big]\; G_3(Q^2), \label{eqn::rho_Gi_basis}
\end{align} 
with a corresponding set of three independent dimensionless form-factors $G_1$, $G_2$, $G_3$. A convenient basis having a clearer physical motivation is provided by the expansion of the vector current in terms of multipoles \cite{Durand:1962zza}, which in this case leads to a set of form-factors,
\begin{align}
G_C &= \left(1 + \tfrac{Q^2}{6m^2} \right) G_1 - \tfrac{Q^2}{6m^2} \, G_2 + \tfrac{Q^2}{6m^2} \left(1 + \tfrac{Q^2}{4m^2} \right) G_3 \nonumber \\
G_M &= G_2 \nonumber \\
G_Q &= G_1 - G_2 + \left(1 + \tfrac{Q^2}{4m^2} \right) G_3, \label{eqn::rho_Gmultipole_basis}
\end{align}
which are proportional to the charge ($C_0$), magnetic dipole ($M_1$), and quadrupole ($C_2$) multipoles respectively\footnote{Note that the relationship between $G_2$ and $G_M$ was presented with a typographic error in \cite{Dudek:2006ej} which is corrected here.}. At $Q^2=0$ they are related to the charge, magnetic moment and quadrupole moment of the vector meson: $G_C(0) = 1$, $G_M(0) = 2 m\cdot  \mu_\rho$, $G_Q(0) = m^2 \cdot Q_\rho$.

The other form-factors we considered above may also be identified with a particular multipolarity -- in the ${\rho \to \pi \gamma}$ transition case the single form-factor is of magnetic dipole ($M_1$) type, while for the $\pi$ cases it is a charge form-factor  $(C_0)$. For stable meson states, time-reversal invariance indicates that the form-factors are real functions of $Q^2$.

\section{Calculation Details \label{sec::calc_details}}

In this first investigation of the extraction of excited state form-factors using distillation, we restrict ourselves to a single ensemble of gauge-field configurations, having three degenerate flavors of dynamical quarks tuned to approximately the physical strange quark mass. This set of anisotropic Clover lattices \cite{Edwards:2008ja, Lin:2008pr} has been used previously in studies of the meson spectrum \cite{Dudek:2009qf,Dudek:2010wm,Dudek:2011tt,Liu:2012ze,Dudek:2013yja}, 
meson decay constants~\cite{Mastropas:2014fsa},
baryon spectrum 
\cite{Edwards:2011jj, Dudek:2012ag, Edwards:2012fx, Padmanath:2013zfa} 
and meson-meson scattering 
\cite{Dudek:2010ew,Dudek:2012gj,Dudek:2012xn,Dudek:2014qha}. 
For the calculations reported on in this paper, we used 535 configurations of lattice volume $(L/a_s)^3 \times (T/a_t) = 16^3\times 128$, with a spatial grid spacing of $a_s \sim 0.12\, \mathrm{fm}$ and a temporal spacing roughly 3.5 times smaller. 

In this calculation we have an exact $SU(3)$ flavor symmetry such that all the octet mesons ($\pi$, $K$, $\eta$) are degenerate with a mass close to $700$ MeV. Where results are expressed in dimensionful units, they are determined from the dimensionless quantities $a_t E$ using the scale-setting procedure,
\begin{equation*}
E = \frac{a_t E}{a_t m_\Omega} \cdot m_\Omega^{\mathrm{phys.}}.
\end{equation*}
where $a_t m_\Omega$ is the $\Omega$ baryon mass calculated on this lattice and $m_\Omega^{\mathrm{phys.}}$ is the experimental value \cite{PDG-2012}.

Our use of a Clover-improved anisotropic quark action introduces an improvement term into the vector current which appears at tree-level. Discussion of the effect of improvement and renormalisation of the vector current will appear in Section \ref{ssec::three_points_improv}.

\section{Two-Point Function Analysis\label{sec::two_points}} 

In order to determine radiative transition amplitudes between meson states within QCD we must first obtain the spectrum of states and find operators, constructed from quark and gluon fields, that reliably interpolate the states of interest from the vacuum. In general a color-singlet operator $\mathcal{O}_i^\dag$ having definite $J^{PC}$ can produce all QCD eigenstates having those quantum numbers,
\begin{equation}
	\mathcal{O}_i^\dag |0\rangle = \sum\nolimits_\mathfrak{n} |\mathfrak{n} \rangle \langle \mathfrak{n} | \mathcal{O}_i^\dag |0\rangle.  \nonumber
\end{equation}
We seek to determine optimized interpolators, $\Omega_\estate{n}^\dagger$, which when acting on the vacuum strongly interpolate only a \emph{single} state with much reduced contributions from other states,
\begin{align*}
\Omega_\estate{n}^\dagger | 0 \rangle &= |\estate{n}\rangle \langle \estate{n} | \Omega^\dagger_\estate{n} | 0 \rangle + \sum_{\estate{m} \neq \estate{n}}  |\estate{m}\rangle \langle \estate{m} | \Omega^\dagger_\estate{n} | 0 \rangle \\
&= |\estate{n}\rangle \langle \estate{n} | \Omega^\dagger_\estate{n} | 0 \rangle + \sum_{\estate{m} \neq \estate{n}}  |\estate{m}\rangle \varepsilon_\estate{m}.
\end{align*}
In essence we seek a procedure by which we can minimize the $\varepsilon_\estate{m}$ ($\estate{m} \neq \estate{n}$) relative to the strength with which our operator creates the $\estate{n}$'th state, $\langle \estate{n} | \Omega^\dagger_\estate{n} | 0 \rangle$. 

We will proceed by using a \emph{basis} of interpolators, $\{ \mathcal{O}_i \}$, to construct two-point correlation functions of the form, 
\begin{equation*}
C_{ij}(t) = \langle 0 | \mathcal{O}^{\,}_i(t) \mathcal{O}_j^\dagger(0) | 0 \rangle,
\end{equation*}
where operators $\mathcal{O}_i$ are color-singlet constructions built from the basic quark and gluon fields of QCD, having the quantum numbers of the desired hadrons. Such correlation functions can be expressed as, 

\begin{equation*}
C_{ij}(t) = \sum_\estate{n} \frac{1}{2E_\estate{n}} \langle 0 | \mathcal{O}_i(0) | \estate{n} \rangle \langle \estate{n} | \mathcal{O}_j^\dagger(0) | 0 \rangle e^{-E_\estate{n} t}, 
\end{equation*}
where the spectrum of eigenstates is seen to control the Euclidean time dependence\footnote{We have introduced our particular choice of state normalization in a finite-volume here - we discuss this in Appendix \ref{app::two-point}. }.

\subsection{Variational Analysis \label{ssec::two_points_variational}}

We propose that within any basis of operators there is a particular linear combination that is most suited to interpolate the lightest state of the spectrum, another linear combination that optimally interpolates the first excited state, a third combination for the second excited state and so on. Thus optimized interpolators take the form $\Omega^\dagger_\estate{n} = \sum_{i}w_i^{(\estate{n})} \mathcal{O}_i^\dagger$, where one can show that the best estimate for the weights $w_i^{(\estate{n})}$, in a variational sense,  comes from solving the generalized eigenvalue problem \cite{Michael:1985ne, Luscher:1990ck, Dudek:2007wv, Blossier:2009kd},
\begin{equation}
C(t) \, v^{(\estate{n})} = \lambda_\estate{n}(t) \, C(t_0)\,  v^{(\estate{n})}. \label{eqn::variational_method}
\end{equation}
Here $C(t)$ is the $N\times N$ matrix whose elements are the correlation functions $C_{ij}(t)$ constructed from the basis of $N$ operators, $\{ \mathcal{O}_i \}$, and $v^{(\estate{n})}$ is a generalized eigenvector. The generalized eigenvalues, or \emph{principal correlators}, ${\lambda_\estate{n}(t)}$ behave like ${e^{-E_\estate{n} (t-t_0)}    }$ at large times, and can be used to determine the spectrum of energy eigenstates. The vectors $ v^{(\estate{n})}$ are orthogonal on a metric, $ v^{(\estate{m})\dagger}C(t_0)  v^{(\estate{n})} = \delta_{\estate{m}\estate{n}}$, where $t_0$ is a reference timeslice. Examination of the orthogonality condition suggests that $t_0$ should be chosen to be sufficiently large such that the correlation functions are dominated by the $N$ lowest-lying states, with heavier states having decayed exponentially to a negligible level. Further considerations on the choice of $t_0$ are presented in \cite{Dudek:2007wv,Blossier:2009kd}.

In practice we solve Eqn~\ref{eqn::variational_method} independently on each timeslice, $t$, so that for each state, $\mathfrak{n}$, we obtain a time series of generalized eigenvectors $v^{(\mathfrak{n})}(t,t_0)$, which we observe to be essentially time-independent with a suitably large choice of $t_0$. In practice we use the mean values (over the ensemble of gauge configurations) of the elements $v^{(\mathfrak{n})}_i$ chosen on a single timeslice to construct the optimized operators as
\begin{equation}
\Omega_\estate{n}^\dagger = \sqrt{2E_\estate{n}} e^{-E_\estate{n}t_0/2} \sum\nolimits_i v^{(\estate{n})}_i \mathcal{O}_i^\dagger, \label{optimized_definition}
\end{equation} 
where the coefficients multiplying the sum are chosen to give the normalization $\langle \estate{n} | \Omega^\dagger_\estate{n} | 0 \rangle = 2E_\estate{n}$, which will prove to be convenient when considering three-point functions.

The procedure of solving the generalized eigenvalue problem in a basis of interpolating fields is carried out independently for each quantum number channel at each possible allowed momentum value.

\subsection{Meson operator construction \label{ssec::two_points_operators}}

A straightforward approach to constructing a basis of operators capable of interpolating meson states is to use fermion bilinears containing some number of spatially directed gauge-covariant derivatives, that is operators of generic structure, 
\begin{equation}
\mathcal{O} \sim \bar{\psi} \Gamma \overleftrightarrow{D} \cdots \overleftrightarrow{D} \psi,  \label{eqn::deriv_ops}
\end{equation}
where $\overleftrightarrow{D} \equiv \overleftarrow{D} - \overrightarrow{D}$. By expressing the vector-like derivatives and Gamma matrices in a circular basis, we can easily construct operators of definite spin using the standard $SO(3)$ Clebsch-Gordon coefficients \cite{Dudek:2009qf,Dudek:2010wm}.  Operators of this type with non-zero momentum can be constructed to have definite helicity, as described in \cite{Thomas:2011rh}. 

In this calculation, QCD is discretised on a grid of points whose spatial structure has a cubic symmetry and as such we have not the complete continuous rotational symmetry of the continuum, but rather a reduced symmetry: the symmetry of the cubic group at rest and the relevant \emph{little group} in a moving frame. A consequence is that instead of having an infinite number of irreducible representations labeled by integer spin $J$ (at rest) we only have access to the finite number of irreducible representations, or \emph{irreps}, of the cube, labelled $A_1$, $T_1$, $T_2$, $E$, and $A_2$.  A corresponding argument applies in-flight, where the continuum helicity labeling is broken down to a finite number of irreps of the little group, see \cite{Thomas:2011rh} for details.

The operators of definite $J$ (or helicity) constructed above can be projected into irreps of the relevant symmetry group using a procedure called \emph{subduction}, 
\begin{equation}
\mathcal{O}^{[J]}_{\Lambda \mu} = \sum_{M=-J}^J \mathcal{S}^{JM}_{\Lambda \mu} \mathcal{O}^{JM}
\end{equation}
where $\Lambda$ labels the cubic irrep and $\mu$ is the `row' of the irrep ($\mu = 1\ldots \mathrm{dim}(\Lambda))$. The subduction coefficients, $\mathcal{S}^{JM}_{\Lambda \mu}$ are tabulated in \cite{Thomas:2011rh,Dudek:2010wm}. These operators have been used extensively to study the excited spectrum of mesons \cite{Dudek:2013yja,Liu:2012ze,Dudek:2011tt,Dudek:2009qf,Dudek:2010wm,Dudek:2012xn,Dudek:2014qha}.

\subsection{Correlator construction through distillation \label{ssec::two_points_corr_const}}

Operators which interpolate hadrons from the vacuum have long been known to do so more effectively if the quark fields are suitably smeared over space \cite{Gusken:1989ad, Allton:1993wc}. An extremely convenient method to do this is provided by \emph{distillation} \cite{Peardon:2009gh}, where the smearing operator is constructed on time-slice $t$ as an outer product of vectors in color and $\vec{x}-$space,
\begin{equation}
\Box_{\vec{x}\vec{y}}(t) = \sum_{n=1}^{N_D} \xi^{(n)}_{\vec{x}}(t)\; \xi^{(n)\dag}_{\vec{y}}(t), \label{distillation}
\end{equation}
where the $N_D$ vectors should be constructed to have strong overlap onto the low-energy quark modes most relevant to low-lying hadron states. A suitable choice for the vectors are the eigenvectors of the gauge-covariant three-dimensional Laplacian on a time-slice ordered by their eigenvalue.

Smearing each quark field, a meson creation operator of fermion bilinear form with momentum $\vec{p}$ is 
\begin{equation*}
\mathcal{O}^\dagger(\vec{p}) = \bar{\psi}_{\vec{x}} \Box_{\vec{x}\vec{y}}\,  e^{-i \vec{p} \cdot \vec{y}}\,
\mathbf{\Gamma}_{\vec{y}\vec{z}} \, \Box_{\vec{z}\vec{w}} \psi_{\vec{w}} 
\end{equation*}
where time, color, and spin indices have been suppressed for brevity and where repeated position indices are summed. The object $\mathbf{\Gamma}$ can be non-local in $\vec{x}$-space, and may for example feature gauge-covariant derivatives as discussed in the previous section.

An advantage of the distillation framework is that it leads to a factorization of two-point functions into matrices describing quark propagation, called \emph{perambulators}, and matrices describing operator construction. A generic connected two-point function using fermion bilinear constructions, in which we explicitly show the smearing operators, can be decomposed as 
\begin{align}
\big\langle 0 \big|& \bar{\psi} \Box \mathbf{\Gamma}^f \Box \psi (t) \; \bar{\psi} \Box \mathbf{\Gamma}^i \Box \psi (0) \big| 0 \big\rangle \nonumber \\
&= - \tau_{nm}(0,t)\, \Phi^f_{mp}(t)\, \tau_{pq}(t,0) \,\Phi^i_{qn}(0) \nonumber
\end{align}
where the perambulators, ${\tau_{pq}(t,0) = \xi^{(p)\dag}(t) \, M^{-1}_{t,0} \, \xi^{(q)}(0)}$, can be obtained by inverting the Dirac matrix $M$ on sources $\{ \xi^{(q)} \}_{q=1\ldots N_D}$ at timeslice 0, and where ${\Phi_{mp}(t)  = \xi^{(m)\dag}(t) \mathbf{\Gamma} \xi^{(p)}(t)}$ encodes the operator construction (here we include the momentum projection in $\mathbf{\Gamma}$).

\subsection{Meson spectra \& optimized operators \label{ssec::two_points_spec}}

\begin{figure*}
\includegraphics[width=0.9\linewidth]{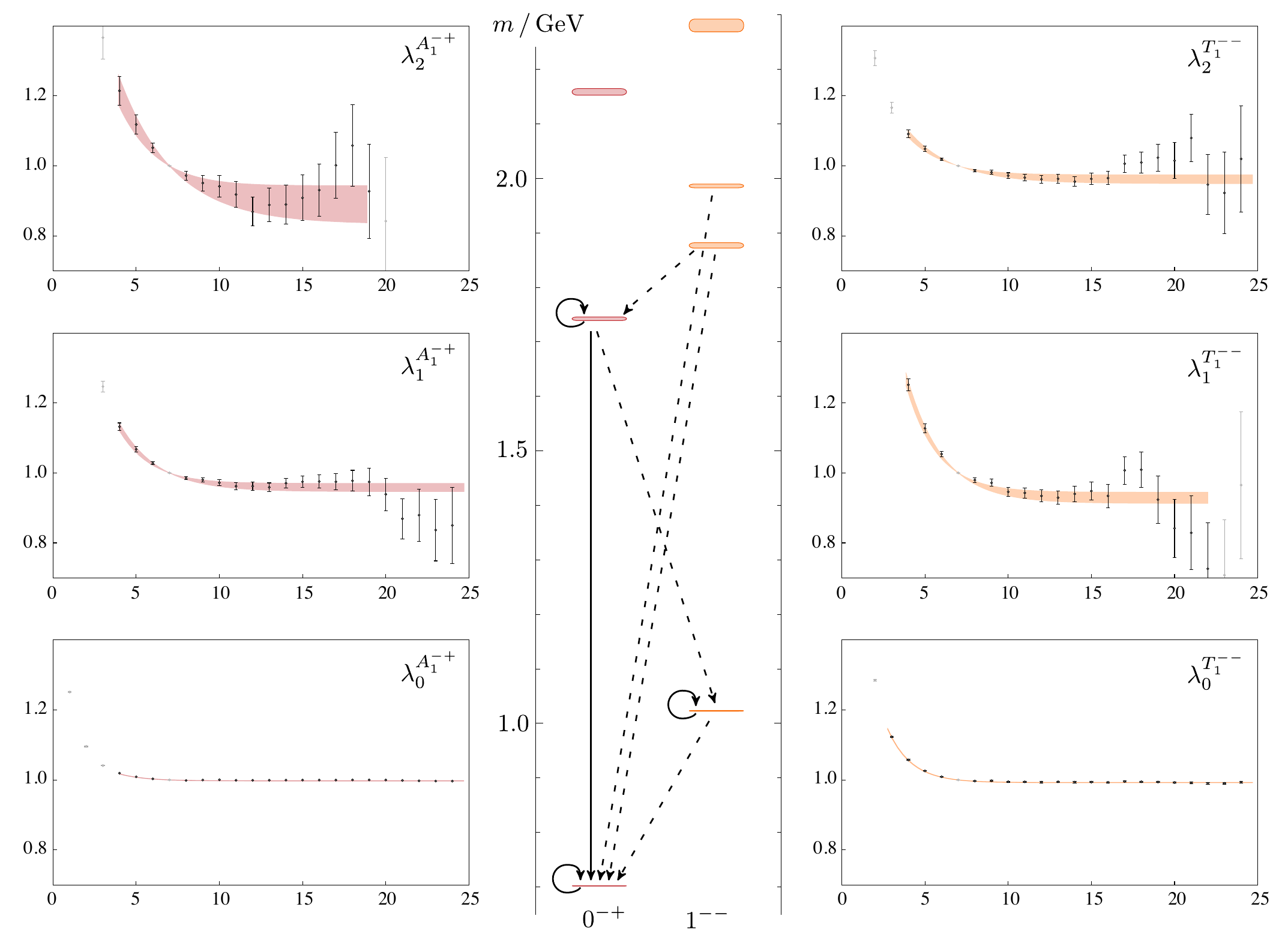}
\caption{
Left(Right) columns: Principal correlators, plotted as $e^{E_\mathfrak{n}(t-t_0)} \lambda_\mathfrak{n}(t)$, for lightest three states in irrep $A_1^{-+}$($T_1^{--}$). Two exponential fits shown, with the resulting mass spectrum of $0^{-+}$ and $1^{--}$ mesons shown in the central column. Radiative transitions and form-factors to be presented in this paper shown by the lines joining the states (solid lines are `charge' transitions, dashed lines are `magnetic' transitions).\label{fig::pcorrs} } 
\end{figure*}

We computed correlation matrices for irreps corresponding to $J^P = 0^-,\, 1^-$ at rest, and to magnitude of helicity $|\lambda| = 0,1$ with nonzero momentum. We made use of $\vec{n}_{\vec{p}} =  [0,0,0],\,[0,0,1],\, [0,1,1],\, [1,1,1], \, [0,0,2]$, where the momentum is expressed in units of $2\pi/L$, $\vec{p} = (2\pi/L)\;\vec{n}_{\vec{p}}$. The quark flavor constructions were chosen to give access to the members of the $SU(3)_F$ octet -- in this case the two-point function Wick contraction contains only a single connected diagram.

In the rest frame we used operator constructions of the type in Eq.~\ref{eqn::deriv_ops} with up to three derivatives while for nonzero momentum we used up to two-derivative constructions. Further details about the construction of derivative operators and the basis used in this calculation can be found in \cite{Thomas:2011rh,Dudek:2009qf,Dudek:2010wm,Dudek:2013yja}. 

The resulting correlation matrices were analyzed using Eq.~\ref{eqn::variational_method}, with each principal correlator, $\lambda_\mathfrak{n}(t)$, being fit with the form,
\begin{equation}
\lambda_\mathfrak{n}(t) = (1- A_\mathfrak{n})\, e^{- E_\mathfrak{n} (t-t_0) } + A_\mathfrak{n}\, e^{- E'_\mathfrak{n} (t-t_0) },
\end{equation}
where the fit parameters are $E_\mathfrak{n}$, $E'_\mathfrak{n}$, and $A_\mathfrak{n}$ -- the second exponential is present to absorb the effect of any states other than $|\mathfrak{n}\rangle$ remaining in the principal correlator. In practice, for suitably large values of $t_0$, we find that the energy scale of $E_\mathfrak{n}'$ is typically at or above the largest energy extracted, $E_{\mathfrak{n} = \mathrm{dim(C)}}$.

\begin{figure}
\includegraphics[width=\linewidth]{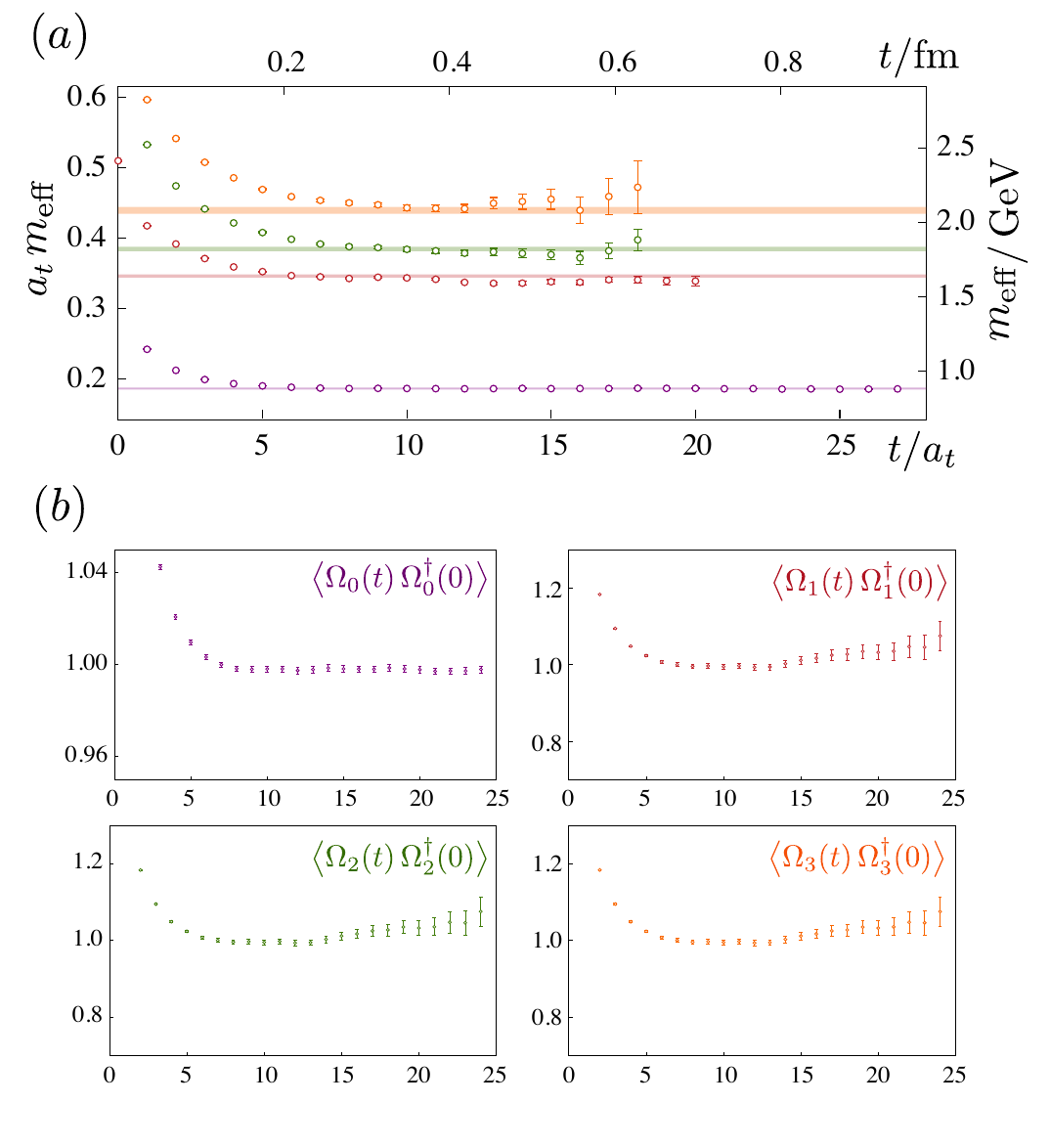}
\caption{
(a) Effective masses of principal correlators for four lightest states in the $A_2^+$ irrep for momentum direction $\vec{n}_{\vec{p}} = [1,0,0]$ along with the energy determined from a two exponential fit. (b) `Optimized' operator correlation functions, $(2E_\mathfrak{n})^{-1} e^{E_\mathfrak{n}t} \big\langle \Omega_\mathfrak{n}(t) \, \Omega_\mathfrak{n}^\dag(0) \big\rangle$, for the four states shown above.
\label{fig::SpecD4A2M}}
\end{figure}

\begin{figure}
\includegraphics[width=0.9 \linewidth]{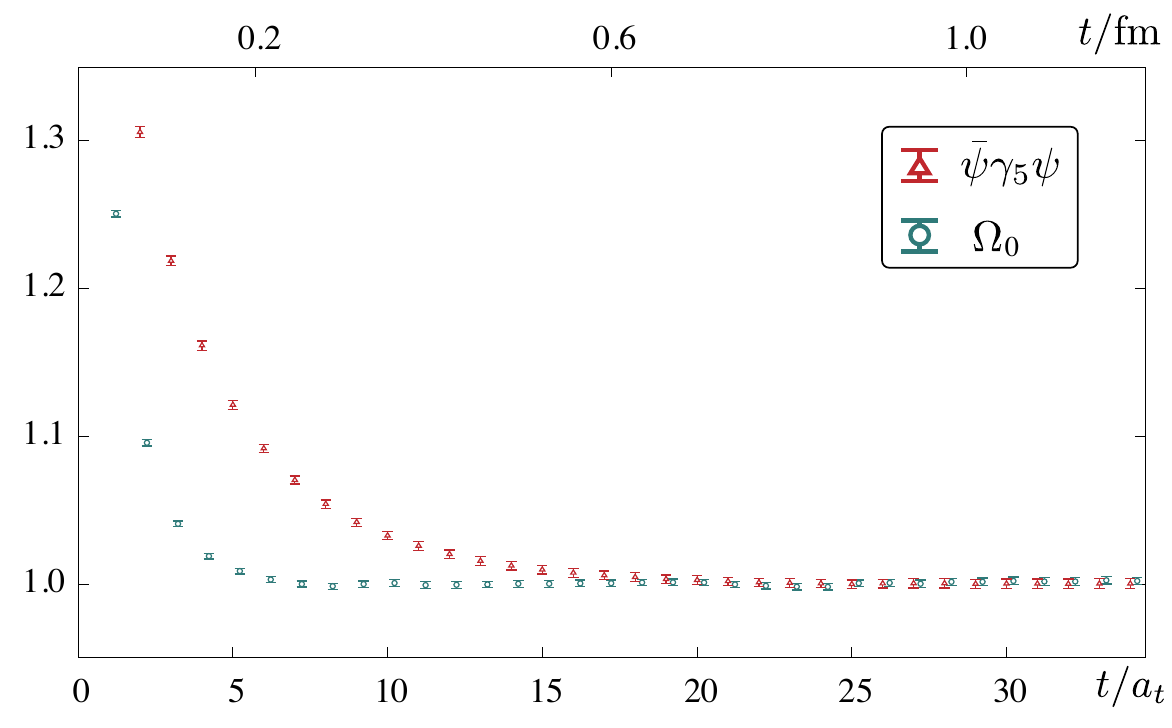}
\caption{The rest-frame pion correlation function using the (distillation smeared) $\bar{\psi}\gamma_5 \psi$ operator (red) versus using the `optimized' operator, $\Omega_0$ (blue). Plotted is $2m_\pi\,  e^{m_\pi t} \, \langle 0| \mathcal{O}(t) \mathcal{O}^\dagger(0) | 0 \rangle  \,/ \, \lvert \langle 0 | \mathcal{O} | \pi \rangle\rvert^2 $.
\label{fig::TwoPointRelaxation}}
\end{figure}

A typical example is presented in Figure \ref{fig::pcorrs}, where the principal correlators for the lightest three states in the $\vec{n}_{\vec{p}}=[0,0,0]$, $\Lambda^{PC} = T_1^{--} (A_1^{-+})$ irreps, which contain $1^{--}(0^{-+})$ mesons, are shown. We make use of a basis of 26 operators in the $T_1^{--}$ channel and 12 operators for $A_1^{-+}$.

Another example is shown in Figure \ref{fig::SpecD4A2M}(a) for the case $\vec{n}_{\vec{p}}=[0,0,1]$, $\Lambda^C = A_2^+$, which contains the helicity zero component of mesons of `unnatural parity' (${J^P=0^-,\, 1^+,\, 2^-,\ldots}$), where we present the \emph{effective mass} of the principal correlator, $m_\mathrm{eff} = \tfrac{1}{\delta t} \log \tfrac{\lambda(t)}{\lambda(t+\delta t)}$ with $\delta t = 3 a_t$. As will be discussed in the next subsection, these states appear to correspond (in order of increasing energy) to the ground-state $\pi$, the $a_1$, the first-excited $\pi'$, and the $\pi_2$.

We note that in both figures we observe that there is negligible curvature present for $t \gtrsim t_0$, indicating that a single state is dominating the correlation function. This observation only holds for sufficiently large choices of $t_0$, and is one guide in the selection of a suitable $t_0$ value. The time beyond which excited state contributions are negligible, which we can call $t^\star$, plays an important role in the construction of our three-point functions in terms of selecting the time separation of the source and sink meson operators. We desire our three-point functions to feature a time region in which the correlation function is dominated by the transition of interest, with contributions from other states having decayed away. In order to achieve this we should separate the source and sink projected operators (Eq.~\ref{optimized_definition}) by at least $t^\star_i + t^\star_f$. It remains possible that the vector current insertion may act to suppress or amplify the contribution of unwanted excited states, but since we do not have this information in advance of the calculation, the above time separation serves as a reasonable estimation of the minimum. 

The source and sink cannot be separated arbitrarily far as the statistical noise on the entire three-point correlation function grows exponentially with increasing separation. We can obtain an estimate of the maximum practical time separation for three-point functions by examining the growth of noise in two-point function principal correlators. As an example in Figure \ref{fig::pcorrs} we see that the ground-state signal remains of high statistical quality out to 25 timeslices (and beyond), while the first and second excited states begin to show significant fluctuations above $t/a_t \sim 15$. Later we will find that while optimized ground-state three-point correlation functions are still statistically precise for time separations as high as 36 timeslices, excited-state correlation functions are not well determined for separations larger than around 20 timeslices.

In practice we solve the matrix problem, Eq.~\ref{eqn::variational_method}, independently on each timeslice (as described in \cite{Dudek:2010wm}). For sufficiently large $t_0$ we find that the elements of the eigenvectors so obtained, $v^{(\mathrm{n})}_i(t)$, are essentially flat for $t \gtrsim t_0$, and in practice we construct our projected operators, Eq.~\ref{optimized_definition}, using ensemble mean values taken from a single timeslice. In Figure \ref{fig::SpecD4A2M}(b) we show the correlation functions $\big\langle 0 \big| \Omega_\mathfrak{n}(t) \Omega^\dag_\mathfrak{n}(0) \big| 0 \big\rangle$, observing that they behave in the manner we expect for optimized operators. The corresponding off-diagonal correlation functions $\big\langle 0 \big| \Omega_\mathfrak{n}(t) \Omega^\dag_\mathfrak{n'}(0) \big| 0 \big\rangle$ for $\mathfrak{n} \neq \mathfrak{n'}$ are statistically compatible with zero for $t \gtrsim t_0$. 

We conclude this section by demonstrating that an optimized ground-state operator constructed as described above, does indeed significantly reduce the contribution of excited states to a correlation function, relaxing to the ground-state more quickly. This can be seen in Figure~\ref{fig::TwoPointRelaxation} where the optimized ground-state operator in $A_1^{-+}$ is compared with the (distillation smeared) $\bar{\psi} \gamma_5 \psi$ operator.

\subsection{Meson dispersion relations \label{ssec::two_points_dispersion}}

\begin{figure}[h]
\includegraphics[width=1.0\linewidth]{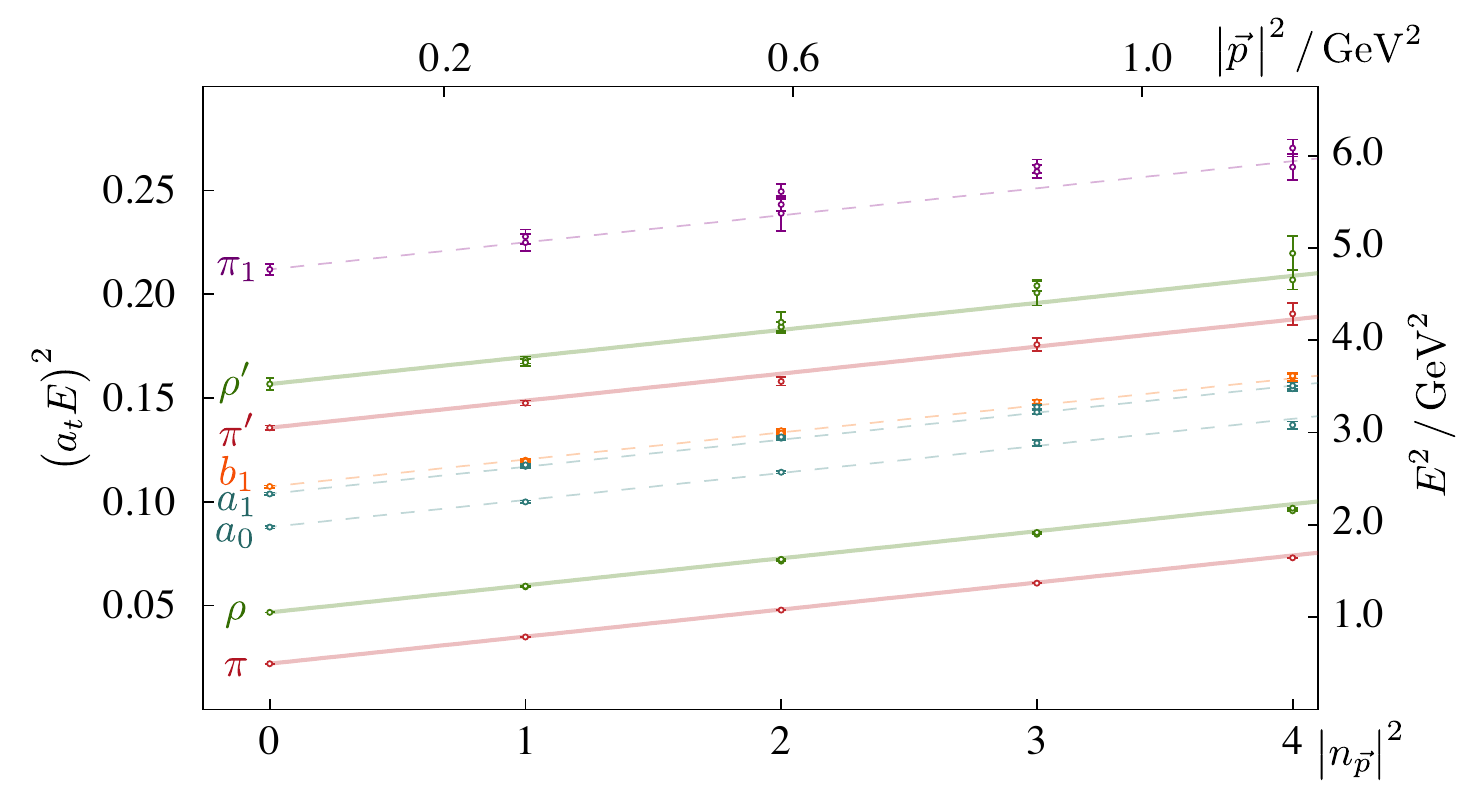}
\caption{Squared energies as a function of $\lvert \vec{n}_{\vec{p}} \rvert^2 =  (\tfrac{L}{2\pi})^2\lvert \vec{p} \rvert^2  $ for selected meson states. The points represent extracted energies for all irreps considered in the variational analysis (e.g. a vector meson like the $\rho$ appears in three irreps with $[1,1,0]$: $A_1, B_1, B_2$), while the lines show the relativistic dispersion relation for an anisotropy $\xi=3.44$. \label{fig::Dispersion}}
\end{figure}

As previously mentioned, we independently compute the energy spectrum of states at each allowed value of $\vec{p}\,$ for each relevant irrep, but we expect to see the same mesons appearing at each momentum with an energy determined by their rest mass and the relevant dispersion relation. After identifying the mesons at each momentum (by their overlap with characteristic operators \cite{Thomas:2011rh}), we may examine their dispersion relation, $E(|\vec{p}\,|)$. This is presented in Figure \ref{fig::Dispersion} for mesons, $\pi, \pi', \rho, \rho'$ and mesons $a_0, a_1, b_1, \pi_1$ (not used in this analysis) where they are all observed to be compatible with the relativistic dispersion relation, $E^2 = m^2 + p^2$, or in temporal lattice units,
\begin{equation*}
\big(a_t E\big)^2 =  \big(a_t m\big)^2 + \left(\tfrac{2\pi}{\xi (L/a_s)}\right)^2 \lvert \vec{n}_{\vec{p}} \rvert^2 ,
\end{equation*}
for a meson of mass $m$, with momentum ${a_t \vec{p} = \tfrac{1}{\xi} a_s \vec{p} = \tfrac{1}{\xi} \tfrac{2\pi}{L/a_s}[n_x, n_y, n_z]}$, with the anisotropy taking the value $\xi = 3.44$. Making use of optimized operators with momenta up to $[0,0,2]$ allows us to sample many values of $Q^2$ in the form-factor extraction.

\section{Three-point functions\label{sec::three-point}}

We now turn to the three-point correlation functions used in this analysis which contain the vector-current matrix elements of interest. Their basic form is
\begin{equation*}
	C_{\mathrm{f} \mu \mathrm{i}}(\Delta t, t) = \big\langle 0 \big| \mathcal{O}^{\,}_\mathrm{f}(\Delta t) \, j_\mu(t) \, \mathcal{O}^\dag_\mathrm{i}(0) \big| 0 \big\rangle,
\end{equation*}
where the operators $\mathcal{O}_{\mathrm{i,f}}$ are capable of interpolating meson states of definite momentum from the vacuum -- a suitable basis was discussed in the previous section. The relation between the correlation function and the desired matrix-element is exposed by inserting complete sets of eigenstates with the quantum numbers of $\mathcal{O}_\mathrm{i}$ and $\mathcal{O}_\mathrm{f}$, and evolving all operators back to the origin of Euclidean time,
\begin{align}
	C_{\mathrm{f} \mu \mathrm{i}}(\Delta t, t) = \sum_{\mathfrak{n}_\mathrm{i},\mathfrak{n}_\mathrm{f} }& 
		\, \frac{1}{2 E_{\mathfrak{n}_\mathrm{f}} } \frac{1}{2 E_{\mathfrak{n}_\mathrm{i}} }
		e^{-E_{\mathfrak{n}_\mathrm{f}}(\Delta t - t)} \,  e^{-E_{\mathfrak{n}_\mathrm{i}}t}  \nonumber \\
	& \times \big\langle 0 \big| \mathcal{O}^{\,}_\mathrm{f}(0) \big| \mathfrak{n}_\mathrm{f} \big\rangle 
	 \big\langle \mathfrak{n}_\mathrm{f} \big| j_\mu(0) \big| \mathfrak{n}_\mathrm{i} \big\rangle 
	\big\langle \mathfrak{n}_\mathrm{i} \big| \mathcal{O}^\dag_\mathrm{i}(0) \big| 0 \big\rangle \nonumber.
\end{align}
The summation runs over all states, but clearly if the separations between the operators are large, $\Delta t \gg t \gg 0$, only the lightest states in the $i$ and $f$ channels will contribute and we can extract the vector-current matrix element between them. However, at modest time separations there will remain subleading exponential contributions from excited states, and these `polluting' terms can be a source of systematic error in the extraction of ground state matrix elements \cite{Aznauryan:2012ba,PhysRevD.83.045010,Horsley:2013ayv,PhysRevLett.96.052001,PhysRevLett.100.171602,Lin:2012ev,Bali:2014gha,Dinter:2011sg,Green:2014xba}. Reducing excited state pollution by simply separating operators by longer Euclidean times is not always practical, due to the increase in statistical noise with increasing separation.

One of our major aims here is to extract excited-state matrix elements, which we may access using the optimized operators described in the previous section. Using the optimized operator for an excited state should lead to a three-point correlation function whose leading behavior at large times is given not by the ground state, but rather by the relevant excited state. If we are interested in the ground-state there is also an advantage to using the appropriate optimized operator in that it will have much reduced overlap onto low-lying excitations (relative to any single operator in the original basis, for example $\bar{\psi} \gamma_5 \psi$, c.f. Figure \ref{fig::TwoPointRelaxation}), leading to a corresponding reduction in the excited state pollution in the three-point correlator.

Three-point correlation functions using optimized operators take the form,
\begin{align}
	C_{\mathfrak{n}_\mathrm{f} \mu \mathfrak{n}_\mathrm{i}}(\Delta t, t) 
		&= \big\langle 0 \big| \Omega^{\,}_{\mathfrak{n}_\mathrm{f}}(\Delta t) \, j^\mu(t) \, \Omega^\dag_{\mathfrak{n}_\mathrm{i}}(0) \big| 0 \big\rangle \nonumber \\
		&= e^{-E_{\mathfrak{n}_\mathrm{f}}(\Delta t - t)} \,  e^{-E_{\mathfrak{n}_\mathrm{i}}t}  \; 
	 	\big\langle \mathfrak{n}_\mathrm{f} \big| j_\mu(0) \big| \mathfrak{n}_\mathrm{i} \big\rangle + \ldots
\end{align}
where the leading time-dependence is that of the states $(| \mathfrak{n}_\mathrm{i} \rangle, | \mathfrak{n}_\mathrm{f} \rangle)$, selected by the choice of optimized operators. The ellipsis represents the residual contributions of other states, which should be significantly suppressed when using optimized operators -- we will explore the degree to which this is manifested in explicit calculation.

\subsection{Correlator construction \& distillation \label{ssec::three_points_corr_const} }

Three-point correlation functions featuring a current insertion require a slight extension of the distillation framework presented in Section \ref{ssec::two_points_corr_const} since the quark fields in the current should be those which appear in the action, which are not smeared. Exposing the smearing operators, the general form required is, 
\begin{equation}
\big\langle 0 \big| \bar{\psi} \Box \mathbf{\Gamma}^f \Box \psi (\Delta t) \cdot  
					\bar{\psi} \mathbf{\Gamma} \psi (t) \cdot
					\bar{\psi} \Box \mathbf{\Gamma}^i \Box \psi (0) \big| 0 \big\rangle, \nonumber
\end{equation}
which, considering for now only the completely connected Wick contraction, can be decomposed as
\begin{align}
\big\langle &0 \big| \bar{\psi} \Box \mathbf{\Gamma}^f \Box \psi (\Delta t) \cdot  
					\bar{\psi} \mathbf{\Gamma} \psi (t) \cdot
					\bar{\psi} \Box \mathbf{\Gamma}^i \Box \psi (0) \big| 0 \big\rangle \nonumber \\
&= - \tau_{nm}(0,\Delta t)\, \Phi^f_{mp}(\Delta t)
 \Big[ \xi^{(p)\dag}(\Delta t) M^{-1}_{\Delta t, t} \mathbf{\Gamma} M^{-1}_{t,0} \xi^{(q)}(0) \Big]
 \Phi^i_{qn}(0) \nonumber \\
 &= - \tau_{nm}(0,\Delta t)\, \Phi^f_{mp}(\Delta t)
\; \mathcal{G}^\mathbf{\Gamma}_{pq}(\Delta t, t, 0) \;
 \Phi^i_{qn}(0), \nonumber
\end{align}
where the object in square brackets, $\mathcal{G}^\mathbf{\Gamma}_{pq}(\Delta t, t, 0)$, is a \emph{generalized perambulator}. It can be obtained through inversion from sources $\{ \xi^{(q)} \}_{q=1\ldots N_D}$ at timeslice 0 to obtain $M^{-1}_{t,0} \xi^{(q)}(0)$, inversion from sources $\{ \xi^{(p)} \}_{p=1\ldots N_D}$ at timeslice $\Delta t$ to obtain $M^{-1}_{t,\Delta t} \xi^{(p)}(\Delta t)$ which can be related to $\xi^{(p)\dag}(\Delta t) M^{-1}_{\Delta t, t}$ using $\gamma_5$ hermiticity, and contraction with the operator insertion, $\mathbf{\Gamma}$, at each value of $t$ between $0$ and $\Delta t$. In this calculation we do not average over multiple time-sources separated by the same value of $\Delta t$, although this could be done to increase statistics.

In our application, exposing the Dirac spin, $\vec{x}-$space and color indices, the current insertion is ${\mathbf{\Gamma}_{\vec{x},\vec{y};a,b}^{\alpha \beta} = \gamma_\mu^{\alpha \beta} \, e^{-i \vec{q} \cdot \vec{x} } \, \delta_{\vec{x},\vec{y}}\,  \delta_{a,b}}$.  In the case of an improved vector current, which we consider later, we require also ${\mathbf{\Gamma}_{\vec{x},\vec{y};a,b}^{\alpha \beta} = \sigma_{4k}^{\alpha \beta} \, e^{-i \vec{q} \cdot \vec{x} } \, \delta_{\vec{x},\vec{y}}\,  \delta_{a,b}}$, yielding another set of generalized perambulators.

\begin{figure}
\includegraphics[width=0.5\linewidth]{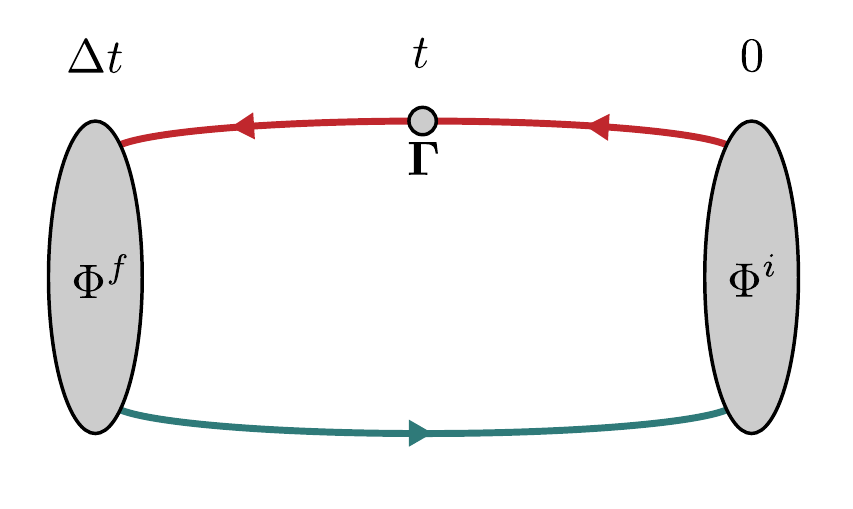}
\caption{Graphical depiction of a connected three-point correlator. The gray blobs represent the source and sink operators carrying the quantum numbers of the initial and final state mesons. The blue line corresponds to a perambulator ($\tau_{pq}(\Delta t,0) = \xi^{(p)\dag}(\Delta t) M^{-1}_{\Delta t,0} \xi^{(q)}(0)$) while the red line represents a \emph{generalized perambulator} as described in the text, $\mathcal{G}^\mathbf{\Gamma}_{pq}(\Delta t, t, 0) = \xi^{(p)\dag}(\Delta t) M^{-1}_{\Delta t, t} \mathbf{\Gamma}^{}_t M^{-1}_{t,0} \xi^{(q)}(0) $, which carries the momentum and quantum numbers of the current insertion. }
\end{figure}

Within this construction we are able to project each operator into definite momentum, and as such we only compute correlation functions in which the momentum is conserved, $\vec{p}_{\mathrm{f}} = \vec{p}_\mathrm{i} + \vec{q}$. Some discussion of momentum conservation in a finite volume appears in Appendix \ref{app::two-point}.

\subsection{Correlation functions using optimized operators \label{ssec::three_points_optimized_ops}}

\begin{figure}[b]
\includegraphics[width=1.03\linewidth]{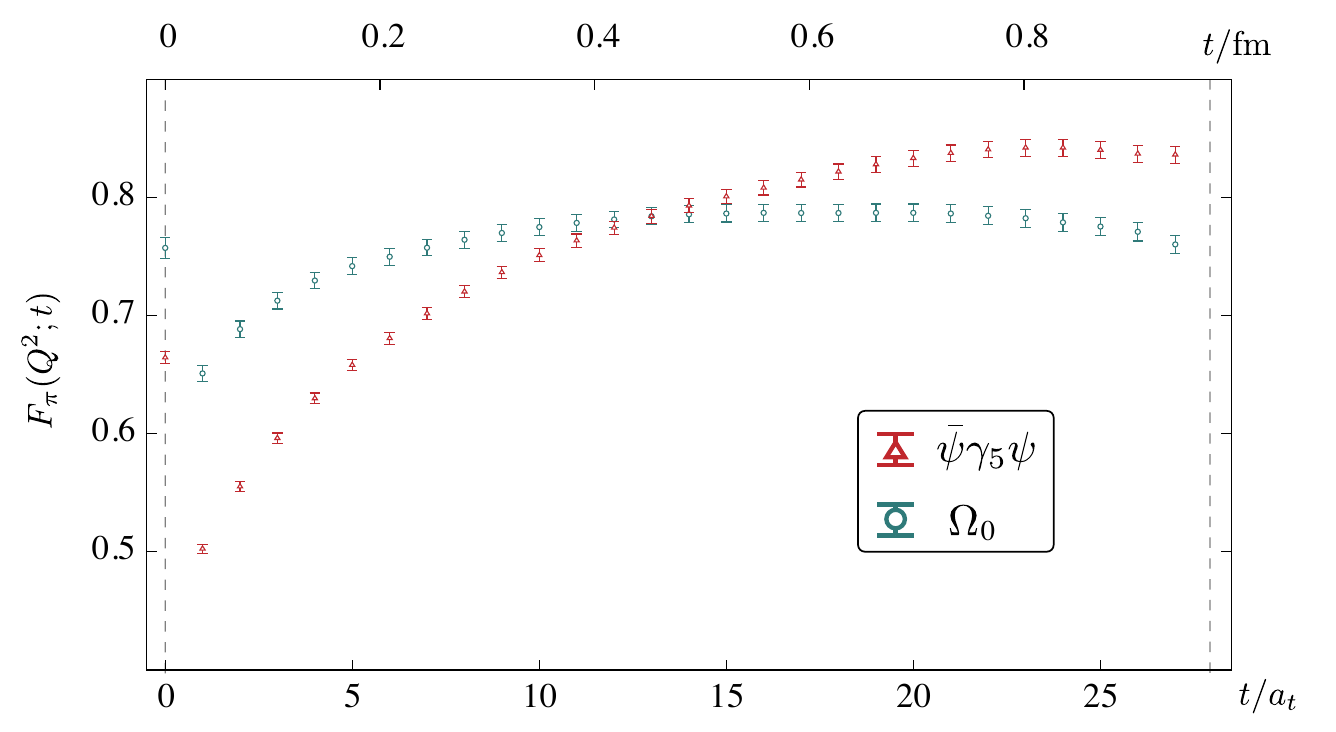}
\caption{Form-factor from vector current three-point function with pion operators at source ($t=0$, $\vec{n}=[0,0,0]$) and sink ($\Delta t = 28\,a_t$, $\vec{n} = [1,0,0]$). Red points correspond to using the `unoptimized' bilinear $\bar{\psi} \gamma_5 \psi$;  the `optimized' operator is shown in blue.
\label{fig::ThreePointRelaxationPion}}
\end{figure}

Turning first to the case of three-point functions with pion-like operators at the source and sink, we plot in Figure~\ref{fig::ThreePointRelaxationPion} the form-factor (as defined in Eqn.~\ref{pion_ff}) extracted from the three-point function, 
\begin{equation*}
\langle 0 | \mathcal{O}_\pi(\Delta t,\vec{p}_f) j^\mu(t, \vec{q}) \mathcal{O}^\dagger_\pi(0,\vec{p}_i) | 0 \rangle
\end{equation*}
where $\mathcal{O}_\pi$ represents either $\bar{\psi}\gamma_5\psi$ (in red) or the optimized operator $\Omega_\pi$ (in blue). The sink operator, located at $\Delta t = 28\, a_t \sim 0.9 \;\mathrm{fm}$, is in the $\Lambda^C = A_2^+$ irrep of momentum $\vec{n}_{\vec{p}_f} = [1,0,0]$, while the source operator, located at $t = 0$, is at rest in the $\Lambda^{PC} = A_1^{-+}$ irrep. We clearly observe that the optimized operators give rise to a signal which is flat over a number of timeslices away from the source and sink, corresponding to the contribution of just the ground-state pion, while the simpler $\bar{\psi}\gamma_5\psi$ operators over this time range always retain a non-negligible pollution from excited states. Such behavior is expected from our two-point function analysis: for example, at rest we find $\left| \tfrac{ \langle 0 | \bar{\psi}\gamma_5 \psi | \mathfrak{n} = 1\rangle}{  \langle 0 | \bar{\psi}\gamma_5 \psi | \mathfrak{n} = 0\rangle} \right| \sim 0.73$, so the distillation smeared operator $\left[ \bar{\psi} \gamma_5 \psi \right]^\dagger $, acting on the vacuum, creates both the ground and first excited state with comparable strength.

\begin{figure}[b]
\includegraphics[width=1.03\linewidth]{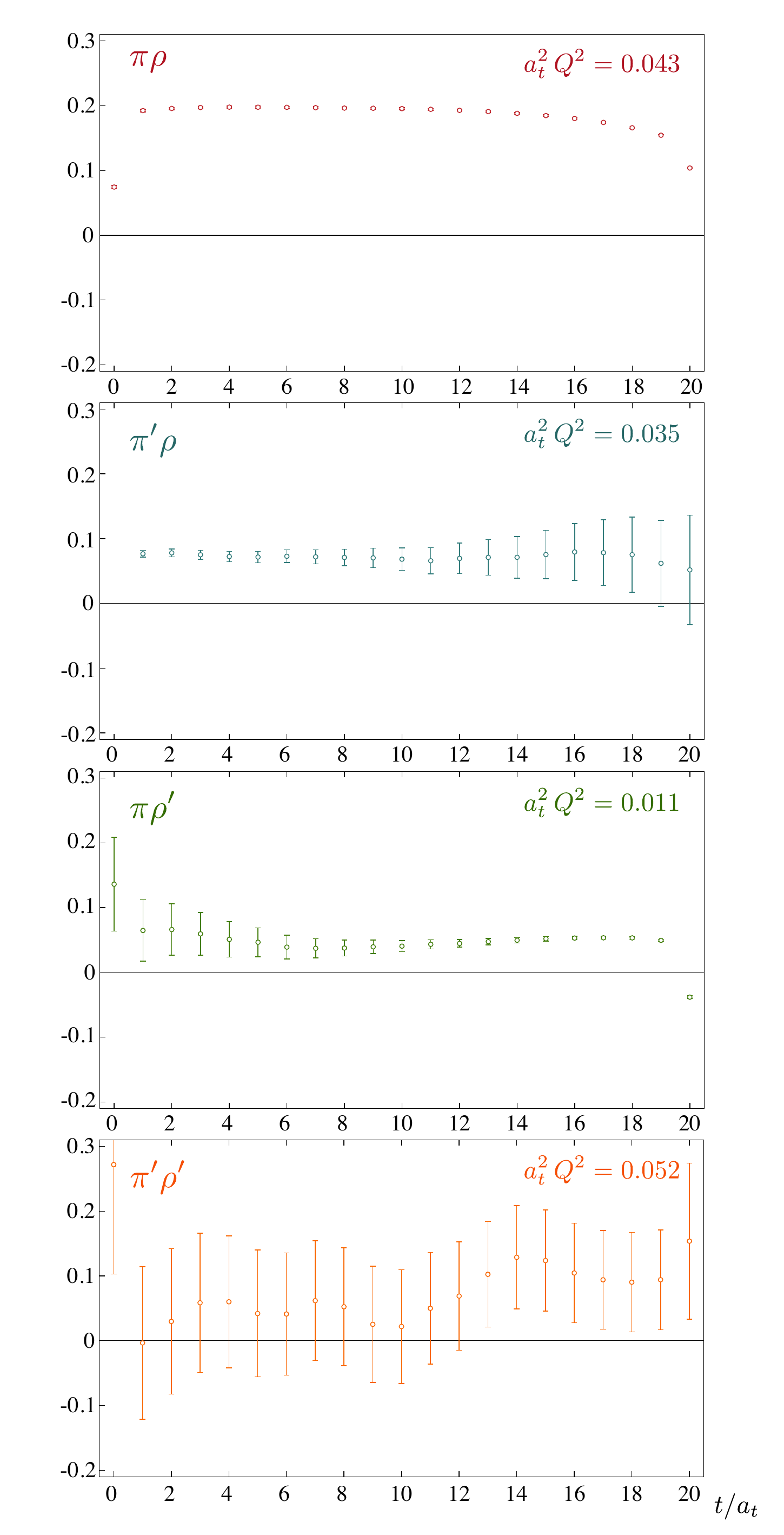}
\caption{ Form-factors extracted from optimized three-point correlation functions with a $\pi$ or $\pi'$ operator with ${\vec{p}=[\text{-}1,0,\text{-}1]}$ at $t=0$ and a $\rho$ or $\rho'$ operator with ${\vec{p}=[1,0,\text{-}1]}$ at $t= 20\,a_t$. The source-sink separation in physical units is roughly 
0.7 fm. \label{rhopiproj}
}
\end{figure}

Our principal motivation for using optimized operators is to get access to transitions involving excited states. In Figure~\ref{rhopiproj} we show matrix elements extracted from three-point correlation functions computed using either the ground-state $\pi$ or first-excited state $\pi'$ optimized operator at the source ($t_i=0,\, \vec{p}_i=[\text{-}1, 0, \text{-}1]$) and either the ground-state $\rho$ or first-excited state $\rho'$ operator at the sink ($t_f = 20\,a_t,\, \vec{p}_f=[1, 0, \text{-}1]$). We observe that there are clear statistically significant signals for excited-state transitions when using the appropriate optimized operators.

\begin{figure*}
\includegraphics[width=0.9\linewidth]{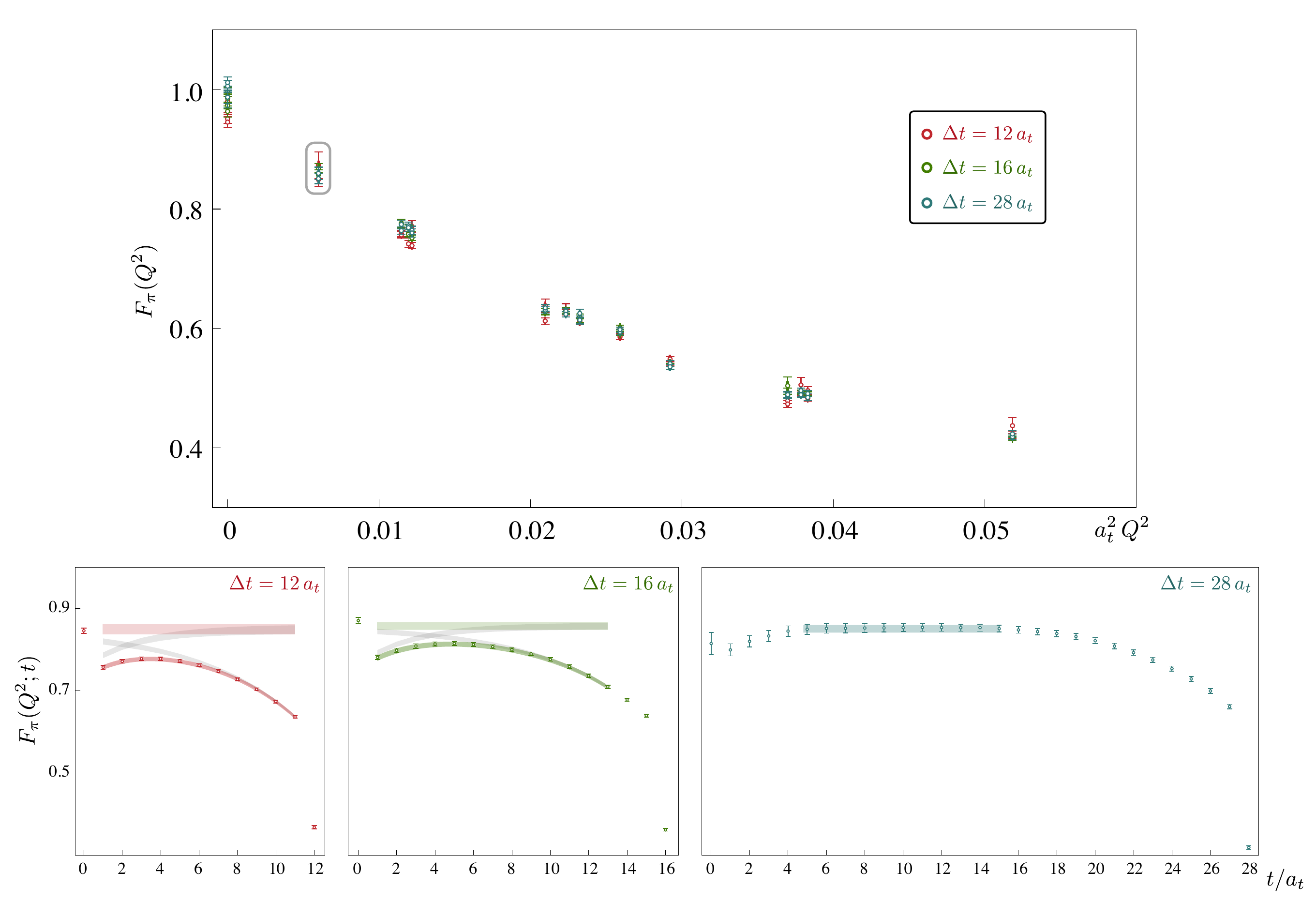}
\caption{ Upper panel shows the pion ground-state form-factor extracted from correlation functions with source-sink separations of $\Delta t = (12,16,28)\, a_t$ using the procedure of fitting with Eqn.~\ref{three_point_fit}. The lower panel illustrates this using the example of $\vec{n}_{\vec{p}_i} = [1,0,0],\, \vec{n}_{\vec{p}_f} = [2,0,0]$.
\label{fig::ThreePointSourceSinkSeparation}  }
\end{figure*}

In general, even for optimized operators, there may still be some residual contamination coming from states that lie beyond the reach of our variational basis, and indeed curvature away from flat behavior as we approach the source or sink timeslice is observed in Figures~\ref{fig::ThreePointRelaxationPion} and \ref{rhopiproj}. 

In order to make maximal use of the time-series data, in particular in those regions where there remains some unwanted excited-state contribution, we opt to perform a correlated fit over a time range with the form,
\begin{equation}
F(Q^2; t) =  F(Q^2)  + f_f \, e^{-\delta E_{f}\,(\Delta t - t)} + f_{i} \,  e^{- \delta E_i\, t} 
\label{three_point_fit}
\end{equation}
where $f_f, \delta E_f, f_i, \delta E_i$ and $F(Q^2)$ are real fit parameters. We make further use only of the constant term, which corresponds to the desired form-factor. Fitting the data to this form also exposes the energy scale of the pollution terms, $\delta E_f$ and $\delta E_i$. Generically, when present, we find that these energies lie at or above the scale of the largest energies we reliably extract in our two-point function variational analysis. In cases where there is a clear extended plateau region, we may exclude the exponential terms and perform a fit to a constant value.

The dependence upon source-sink separation, $\Delta t$, for the ground-state pion form-factor can be seen in Figure~\ref{fig::ThreePointSourceSinkSeparation}. The lower panel shows the illustrative case ${\vec{n}_{\vec{p}_i} = [1,0,0]}$, ${\vec{n}_{\vec{p}_f} = [2,0,0]}$, where we observe that there is only a visible plateau region for $\Delta t = 28\, a_t$, while for the shorter separations, $\Delta t = 12\, a_t, 16\, a_t$, we make use of a fit using Eqn.~\ref{three_point_fit}. The resulting values of $F_\pi(Q^2)$ are observed to be compatible -- other time separations were also explored with similar results. The upper panel shows that this procedure is generally applicable and leads to form-factors from each time-separation that are in agreement across a range of $Q^2$.

In practice, while we extract a very large number of form-factor determinations at many $Q^2$-values, we choose to make use of only those where application of Eq.~\ref{three_point_fit} to $F(Q^2;t)$ shows modest excited-state contributions. Any cases where a clear trend toward a constant value is not visible are discarded.

\subsection{Extracting multiple form-factors \label{sec::3pt_lin_sys}}

Eqn. \ref{decomp} presents the general form of the decomposition of a vector-current matrix element into independent form-factors, $F_i(Q^2)$, and the corresponding kinematical factors, $K_i$, which depend upon momenta and helicities. Moving to a more complete notation including momentum and helicity labels, our three-point correlation functions may be written, 
\begin{align}
&\!\!\big\langle 0 \big| \Omega^{\,}_{\mathfrak{n}_\mathrm{f},\vec{p}_\mathrm{f}, \lambda_\mathrm{f}}(\Delta t) \, j^\mu_{\vec{q}}(t) \, \Omega^\dag_{\mathfrak{n}_\mathrm{i},\vec{p}_\mathrm{i}, \lambda_\mathrm{i}}(0) \big| 0 \big\rangle \nonumber \\
&\!\!= e^{-E_{\mathfrak{n}_\mathrm{f}}(\Delta t - t)}  e^{-E_{\mathfrak{n}_\mathrm{i}}t}   
	 	\big\langle \mathfrak{n}_\mathrm{f}, \vec{p}_\mathrm{f}, \lambda_\mathrm{f} \big| j^\mu(0) \big| \mathfrak{n}_\mathrm{i},\vec{p}_\mathrm{i}, \lambda_\mathrm{i} \big\rangle + \ldots \nonumber \\
		&\!\!= e^{-E_{\mathfrak{n}_\mathrm{f}}(\Delta t - t)}  e^{-E_{\mathfrak{n}_\mathrm{i}}t} 
			  \!\sum_i \! K^\mu_i\big(\mathfrak{n}_\mathrm{f}, \vec{p}_\mathrm{f}, \lambda_\mathrm{f} ; \mathfrak{n}_\mathrm{i}, \vec{p}_\mathrm{i}, \lambda_\mathrm{i} \big) \, F_i(Q^2) + \ldots, \nonumber
\end{align}
where as previously the ellipsis represents possible pollution from states other than $(| \mathfrak{n}_\mathrm{i} \rangle, | \mathfrak{n}_\mathrm{f} \rangle)$, which will have residual time-dependence, but which as shown in the previous section are suppressed when using optimized operators.

Any one correlator provides, in general, an underdetermined linear system for the multiple form-factors we wish to extract. By using different combinations of initial and final state helicities and momenta all at the same $Q^2$, we can build a linear system which is constrained or over-constrained, from which we can determine the set $\{ F_i(Q^2) \}$. Removing the known Euclidean time-dependence, $e^{-E_{\mathfrak{n}_\mathrm{f}}(\Delta t - t)} \,  e^{-E_{\mathfrak{n}_\mathrm{i}}t} $, from the correlation functions, we should be left with objects which are time-independent up to pollution from other states, which should be modest for optimized operators. For fixed state choices, $\mathfrak{n}_\mathrm{i}, \, \mathfrak{n}_\mathrm{f}$, using an indexing $a = \big(\vec{p}_\mathrm{i}, \lambda_\mathrm{i}; \mu; \vec{p}_\mathrm{f}, \lambda_\mathrm{f}\big)$, we can write the linear system,
\begin{align}
&\Gamma_a(\Delta t, t) \nonumber 
\\ &\equiv e^{E_{\mathfrak{n}_\mathrm{f}}(\Delta t - t)} \,  e^{E_{\mathfrak{n}_\mathrm{i}}t} \, \big\langle 0 \big| \Omega^{\,}_{\mathfrak{n}_\mathrm{f},\vec{p}_\mathrm{f}, \lambda_\mathrm{f}}(\Delta t) \, j^\mu_{\vec{q}}(t) \, \Omega^\dag_{\mathfrak{n}_\mathrm{i},\vec{p}_\mathrm{i}, \lambda_\mathrm{i}}(0) \big| 0 \big\rangle\nonumber \\
 &= \sum\nolimits_i K_i\big(a \big) \, F_i(Q^2) + \ldots, \label{eqn::linear_system}
\end{align}
which is of the form $\mathbb{\Gamma} = \mathbb{K} \cdot \mathbb{F}$ with $\mathbb{\Gamma}$ a vector over $a$, $\mathbb{K}$ a rectangular matrix with indices $a,i$, and $\mathbb{F}$ a vector of form-factors, indexed by $i$. This may be converted into a system featuring a square matrix: $ \mathbb{K}^\dag \mathbb{\Gamma} = \Big[ \mathbb{K}^\dag \mathbb{K} \Big] \mathbb{F}$, which can be inverted using SVD. In the case where only a single form-factor contributes this procedure can still be followed as a way to average over rotationally equivalent momentum combinations.

In practice we solve this system independently for each value of $t$ between $0$ and $\Delta t$ -- if our optimized operators were perfect, leading to no pollution from other states, we would obtain the same form-factors on each timeslice. In fact we obtain $F_i(Q^2; t)$, where the time-dependence is fitted as described in the previous section to account for the presence of pollution from other states. 

This approach was previously used in \cite{Dudek:2006ej} and \cite{PhysRevD.79.094504} in the extraction of charmonium form-factors.

\subsection{Cubic symmetry \label{ssec::three_points_cub_sym}}

\begin{figure*}
\includegraphics[width=\linewidth]{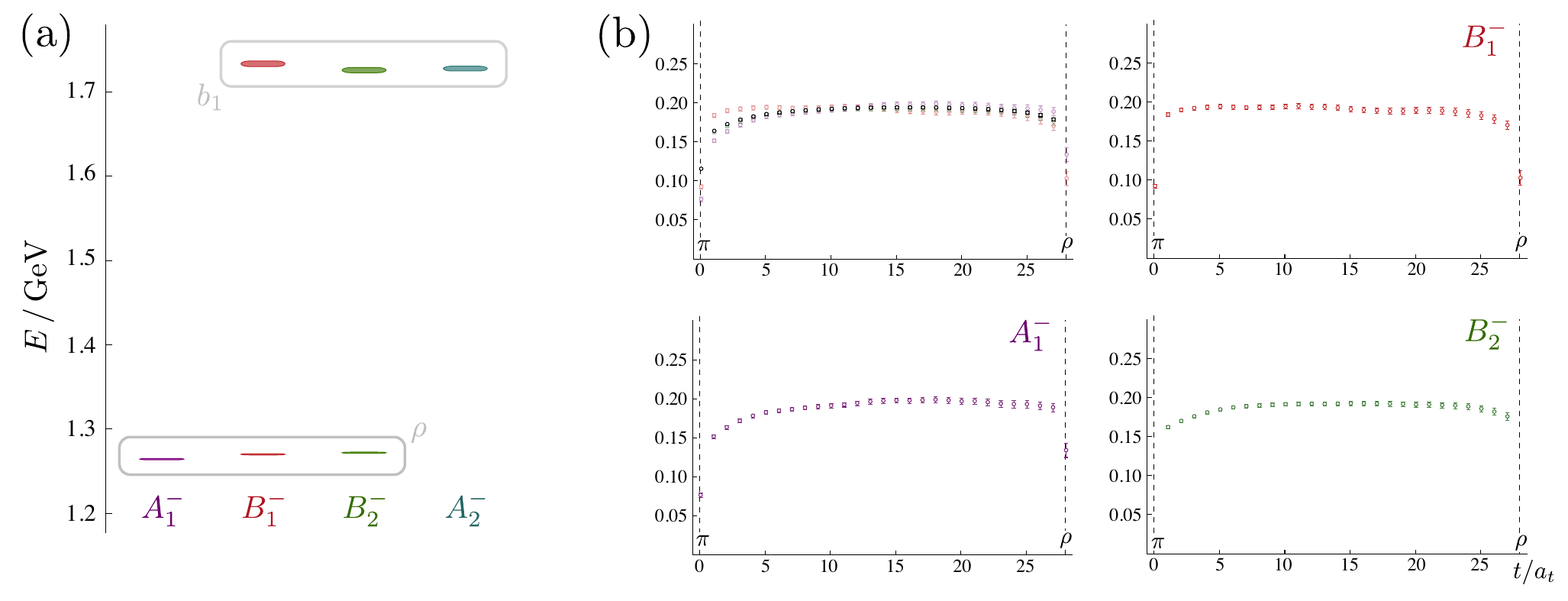}
\caption{
(a) The energy spectrum for the little-group irreps, $\Lambda^C = A_1^-,B_1^-,B_2^-,A_2^-$  with momentum $\vec{n}_{\vec{p}} = [1,1,0]$, which are observed to contain states with degeneracy pattern consistent with the lowest $\rho(1^{--})$ meson and the lowest $b_1(1^{+-})$. (b) The transition form-factors extracted from the correlation functions $\langle 0 | \Omega^{(\Lambda)}_{\rho}(28,\vec{p}\,') \, j^\mu(t,\vec{q}) \, \Omega_\pi^\dagger(0,\vec{p}\,) | 0 \rangle$ with $\vec{n}_{\vec{p}\,'} = [0,1,1]$, $\vec{n}_{\vec{p}} = [0,\text{-1},1]$ for the irreps $\Lambda^C = A_1^-,B_1^-,B_2^-$, which we observe to have consistent values, differing only in the amount of excited state pollution (the current is projected into the $E_2$ irrep of momentum direction  $\vec{n}_{\vec{q}} = [0,2,0]$). The black points in the upper left are the result of solving the linear system described in Section~\ref{sec::3pt_lin_sys}, including all equivalent rotations of the source and sink momenta.
\label{fig::restoration_of_rotational_symmetry}}
\end{figure*}

A consequence of discretizing QCD on a hypercubic grid is that the theory does not possess the full three dimensional rotational symmetry of the continuum. Instead, we are restricted to a subset of rotations that leave the cube invariant. This smaller symmetry group has only a finite number of irreducible representations into which the infinite set of continuum representations labelled by integer spin, $J$, must be \emph{subduced}. A simple example is $J=2$, where the five equivalent `rows' ($M=-2 \ldots 2$) get distributed into a three-dimensional irrep called $T_2$ and a two-dimensional irrep called $E$. Because there are only a finite number of these irreps, they must accommodate multiple values of $J$, such that $T_2$ also contains parts of $J=3,4\ldots$. For systems with non-zero momentum, the symmetry group is called the `little group', and the corresponding subduction is from helicity, $\lambda$. Tables of the spin/helicity content of cubic irreps can be found in \cite{Thomas:2011rh}.

To correctly reflect the symmetry of our theory then, we should label our correlation functions according to irreducible representations of the cubic symmetry. In practice this is what we do by computing using the subduced operators introduced in Section \ref{ssec::two_points_operators}. Using these operators, the three-point functions take the form,
\begin{equation}
	\big\langle 0 \big| \Omega_{\mathfrak{n}_\mathrm{f}, \vec{p}_\mathrm{f}}^{\Lambda_\mathrm{f},\mu_\mathrm{f}}(\Delta t) \, j_{\vec{q}}^{\Lambda_\gamma, \mu_\gamma}(t) \, \Omega_{\mathfrak{n}_\mathrm{i}, \vec{p}_\mathrm{i}}^{\Lambda_\mathrm{i},\mu_\mathrm{i}\dag}(0) \big| 0 \big\rangle, \label{subduced_3pt}
\end{equation}
where the indices $\Lambda$, $\mu$ label the cubic group irrep and the `row' ($1\cdots\text{dim}(\Lambda)$) of the irrep. 

Considering only the cubic symmetry of the lattice, and not any underlying continuum-like symmetry, we would not expect there to be any relationship between different irreps. Furthermore matrix element decompositions should be defined in terms of the irreps of the cube, not in terms of hadrons of definite spin. For example a correlation function with a $T_2$ operator at the source should take values that need not be related to one with an $E$ operator at the source.

However, were there really to be no relation, we could hardly claim to be approximating QCD in a realistic manner. In practical calculations it should be the case that through a combination of sufficiently fine lattice spacing, reduction of discretization artifacts through improvement of the action \cite{Symanzik:1983dc}, and interpolation of hadrons using operators smoothed over many lattice sites \cite{Davoudi:2012ya}, that the continuum symmetry is manifested to a good approximation with only small deviations. For example we might expect to see a relation between $T_2$ and $E$ correlation functions corresponding to them originating from the same $J=2$ meson. In previous two-point function calculations we have observed that the rotational symmetry of the continuum theory is clearly visible in relations amongst the irreps both for eigenstate masses and the values of matrix elements $\big\langle 0 \big| \mathcal{O}_{\Lambda \mu}^{[J]} \big| \mathfrak{n} \big\rangle$ \cite{Dudek:2009qf, Dudek:2010wm}.  

Since we expect to see a comparable restoration of the rotational symmetry in this calculation, we do not attempt to build decompositions according to the symmetries of the cube, rather making use of the continuum-like helicity decompositions presented earlier, subduced  into irreducible representations of the cube. 

A slight additional complication in this analysis arises from our use of anisotropic gauge configurations in which the space and time directions are discretized with different spacings. Spatially directed currents will need to be renormalized separately from temporal currents and the discretization effects along the two directions are expected to be different -- in explicit calculation we will not mix spatially directed currents with their temporal counterparts. Had we used isotropic lattices the temporal component of the vector current would be related to the spatial components, however here we will keep them separate with the temporal component of the current subducing differently from the spatial components. For spatial components\footnote{Here we have made the choice to treat the vector current as a creation operator as opposed to an annihilation operator. The alternate definition would induce changes of phases throughout the calculation (e.g. subduction coefficients and momentum projection).}, the subduced current is ${j_{\vec{q}}^{\Lambda_\gamma, \mu_\gamma} = \sum_\lambda \left[ \mathcal{S}^{\Lambda_\gamma, \mu_\gamma}_{J=1,\lambda} \right]^* \, j^\lambda}$ where ${j^\lambda = \vec{\epsilon}(\vec{q}, \lambda) \cdot \vec{j}}$, whereas temporal components subduce as ${j_{\vec{q}}^{\Lambda_\gamma, \mu_\gamma} = \left[ \mathcal{S}^{\Lambda_\gamma, \mu_\gamma}_{J=0}\right]^* \, j^{\nu=0}}$ .

In order to relate the irrep-based correlation functions that we compute, Eq.~\ref{subduced_3pt}, to the helicity-based decompositions presented in Eq.~\ref{decomp}, we define subduced matrix elements, which for the spatial current case take the form,
\begin{align}
\big\langle  \mathfrak{n}_\mathrm{f}, &\vec{p}_\mathrm{f}, \Lambda_\mathrm{f}, \mu_\mathrm{f} \big| j^{\Lambda_\gamma, \mu_\gamma} \big| \mathfrak{n}_\mathrm{i}, \vec{p}_\mathrm{i}, \Lambda_\mathrm{i}, \mu_\mathrm{i} \big\rangle  \nonumber \\
	= &\sum_{\lambda_\mathrm{i}, \lambda_\gamma, \lambda_\mathrm{i}} S^{\Lambda_f, \mu_f}_{J_f,\lambda_f}  \left[S^{\Lambda_\gamma ,\mu_\gamma}_{J_\gamma =1 ,\lambda_\gamma} \right]^* \left[S^{\Lambda_i ,\mu_i}_{J_i,\lambda_i}\right]^*  \label{eqn::subduced_decomp} \\
	 &\times \sum\nolimits_l \, \vec{\epsilon}(\vec{q},\lambda_\gamma) \cdot \vec{K}_l\big(h_{f,J_f}\big(\lambda_f,\vec{p}_f); h_{i,J_i}(\lambda_i,\vec{p}_i) \big) \, F_l(Q^2). \nonumber
\end{align}
A similar expression exists for the temporal portion of the current, here it will subduce into a ``scalar'' one dimensional irrep.  

The use of the Lorentz-covariant decomposition in this expression implies relationships between different irreps that we must establish are present in the computed correlation functions for this approach to be considered reasonable. 

In Figure~\ref{fig::restoration_of_rotational_symmetry}(a) we show an example of the extracted spectrum across little-group irreps, ${\Lambda^C = A_1^-, B_1^-, B_2^-, A_2^-}$ for ${n_{\vec{p}}=[1,1,0]}$, where the distribution of states matches the expected subduction patterns for a pair of meson states, a lighter $\rho$ ($J^{PC}=1^{--}$) state and a heavier $b_1$ ($J^{PC}=1^{+-}$) state. Forming the optimized operator for the $\rho$ state in each of the $A_1^-, B_1^-, B_2^-$ irreps, we can compute the three-point function, 
\begin{equation*}
\langle 0 | \Omega_\rho(\Delta t,\vec{p}\,')\,  j^\mu(t,\vec{q}\,) \,  \Omega_\pi^\dagger(0,\vec{p}\,) | 0 \rangle,
\end{equation*}
for $\vec{n}_{\vec{p}\,'} = [0,1,1]$, $\vec{n}_{\vec{p}} = [0,\text{-}1,1]$, and $\vec{n}_{\vec{q}} = [0,2,0]$,  across the little-group irreps $A_1^-, B_1^-, B_2^-$ at the sink. The source operator is the optimized operator for the ground-state pion in the $A_2^+$ irrep. The three different sink irrep choices correspond to the subduced versions of the three helicity projections of a vector meson. For $\Delta t = 28\,  a_t$, the resulting form-factor is plotted in Figure~\ref{fig::restoration_of_rotational_symmetry}(b), where we observe that while the amount of excited state pollution differs slightly in each irrep, the form-factor values are consistent, indicating that we are observing components of the same $1^{--}$ meson in the three irreps.

In general we find that the relationships between irreps implied by using Lorentz-covariant decompositions of matrix elements of hadrons of definite spin as in Eq.~\ref{eqn::subduced_decomp} are present in our correlation functions and that the corresponding linear systems of the type expressed in Eq.~\ref{eqn::linear_system} can be solved satisfactorily.  

The cubic nature of the \emph{boundary} of the lattice has additional implications which impact the properties of unstable hadron states -- we defer a discussion of this point to Section \ref{sec::summary}.

\subsection{Renormalization \& improvement of the vector current \label{ssec::three_points_improv}}

The local vector current $\bar{\psi}\gamma^\mu \psi$ is not conserved with a Clover discretized fermion action, and should be renormalized multiplicatively by a factor $Z_V$. We determine this factor non-perturbatively by computing the charge form-factor of the $\pi^+$ or $\rho^+$ meson at $Q^2=0$, where the continuum value corresponds to the charge of the meson in units of $e$, $F(0) = 1$, so that 
\begin{equation}
	Z_V = \frac{F_\pi^\mathrm{cont.}(0)}{F^\mathrm{lat.}_\pi(0)} = \frac{1}{F^\mathrm{lat.}_\pi(0)} .
\end{equation}

\begin{figure}[b]
\includegraphics[width=\linewidth]{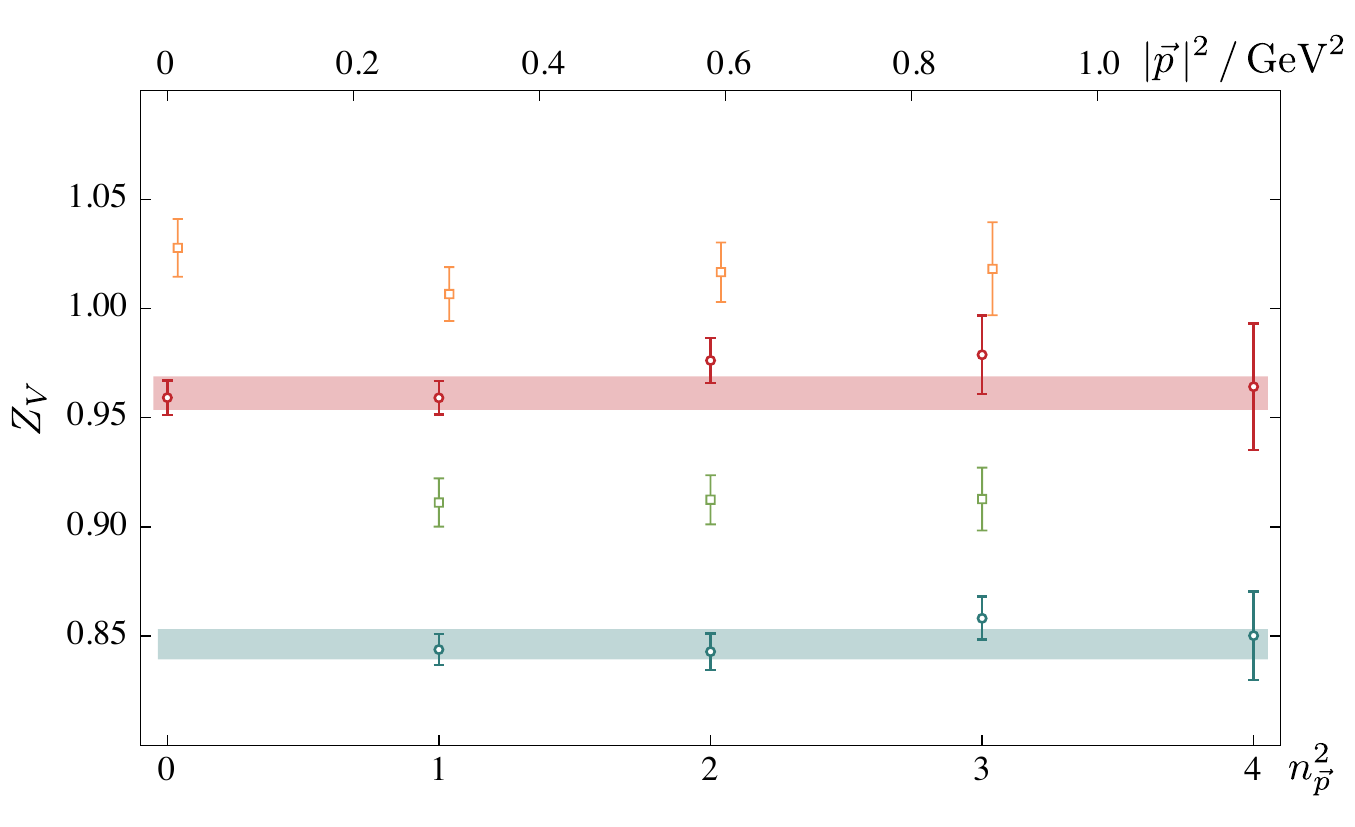}
\caption{Vector current renormalization factor extracted from $Q^2=0$ form-factors of the pion (circles) and the $\rho$ (squares). Spatial current in blue, green, temporal current in red, orange. \label{fig::Z_V}}
\end{figure}

On an anisotropic lattice, where we have treated space and time differently in the action, there can be one $Z_V$ for the spatial vector current, $\bar{\psi}\gamma^i \psi$, and another for the temporal vector current, $\bar{\psi}\gamma^0 \psi$. We extract the zero momentum transfer form-factor from correlation functions with identical source and sink momentum, ${\langle 0 | \Omega_\pi(\Delta t, \vec{p}) \, j^\mu(t, \vec{q}=\vec{0}) \, \Omega^\dag_\pi(0, \vec{p}) | 0 \rangle}$. In Figure \ref{fig::Z_V} we show our extracted values of $Z_V$ observing no significant dependence on the momentum $\vec{p}$. We do however observe some dependence upon whether we extract from the pion form-factor or from the $\rho$ form-factor. The dependence upon the state, a discretization effect, is expected as we have not used a conserved current. In addition, some of the discrepancy can be attributed to imperfect tuning of the anisotropy parameters in the fermion and gauge action~\cite{Edwards:2008ja}, where the measured fermion anisotropy is 3.44 compared to the target value of 3.5.

Using the $\pi$ extraction, which is statistically most precise, performing a correlated average over momenta, we obtain,
\begin{equation}
	Z_V^s = 0.846(6), \;\; Z_V^t = 0.961(7), \label{ZV}
\end{equation}
for the spatial and temporal renormalization factors respectively. All subsequent presentations of form-factor values in this paper have been multiplicatively renormalized by the ensemble mean value of the relevant choice of these two factors\footnote{the statistical uncertainty on $Z_V$ can be considered an overall systematic error on the normalization of form-factors.}.

The anisotropic discretization also introduces a tree level $\mathcal{O}(a)$ improvement term not present in isotropic calculations which amounts to replacing the Euclidean current ${j_\mu =\bar{\psi}\gamma_\mu\psi }$ with 
\begin{align}
j_4 &=  \bar{\psi}\gamma_4 \psi + \tfrac{1}{4} \tfrac{\nu_s}{\xi} ( 1- \xi) \, a_s \partial_j \big( \bar{\psi}\sigma_{4j} \psi\big)  \nonumber \\
j_k &=  \bar{\psi}\gamma_k \psi + \tfrac{1}{4} ( 1 - \xi) \, a_t \partial_4 \big( \bar{\psi}\sigma_{4k} \psi\big),
\label{improvement}
\end{align}
where $\xi = a_s/a_t = 3.44$ is the anisotropy, and $\nu_s = 1.3$ is a parameter appearing in the anisotropic fermion action \cite{Edwards:2008ja, Chen:2000ej}. Details of the derivation of the improved current are deferred to Appendix \ref{app::Improv}.

\begin{figure}[b]
\includegraphics[width=\linewidth]{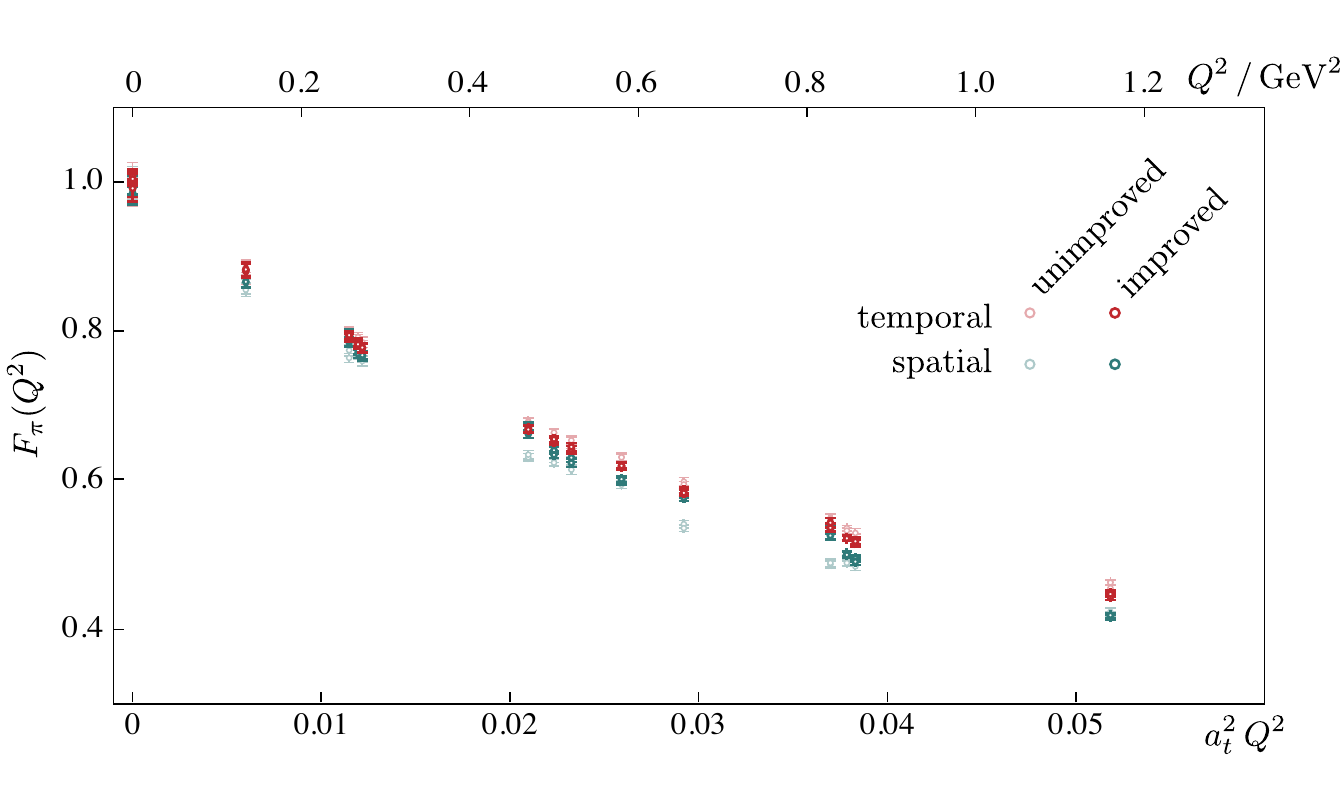}
\caption{The pion ground-state form-factor for unimproved ($\bar{\psi} \gamma_\mu \psi$) and improved (Eqn.~\ref{improvement}) currents.
\label{fig::pi_improvement}}
\end{figure}

In Figure \ref{fig::pi_improvement} we plot our determination of the pion form-factor at a range of $Q^2$ values using both the unimproved and the improved current. We note that the addition of the improvement term, which over the $Q^2$ range considered provides only a small shift, does bring the spatial and temporal current extractions into better agreement. Since we observe the effect of the improvement term to be small, and our main aim is to explore the use of `optimized' meson operators, in the remainder of this report, with the exception of the ground-state pion form-factor, we will make use of only the spatial component of the unimproved vector current.

\pagebreak

\section{Extracted form-factors \& transitions\label{sec::results}}

In this section we present form-factors and transitions for the lightest few isovector pseudoscalar and vector mesons. We make use of the current ${  j^\nu = +\frac{2}{3}\bar{u}\gamma^\nu u -\frac{1}{3}\bar{d}\gamma^\nu d -\frac{1}{3}\bar{s}\gamma^\nu s  }$, such that the form-factors are in units of $e$, the magnitude of the electron charge. This calculation is performed with three flavors of dynamical quark all having the same mass, tuned approximately to the physical strange quark mass. We extract vector current matrix elements between ${(I,I_z) = (1,+1)}$ members of $SU(3)_F$ octets. Disconnected diagrams do not contribute to the amplitudes considered in this analysis as demonstrated in Appendix \ref{app::flavour}, where the flavor structure of the current is explored further.

\subsection{Form-factors}

\subsubsection{$\pi$ form-factor}

\begin{figure}[h]
  \includegraphics[width=\linewidth]{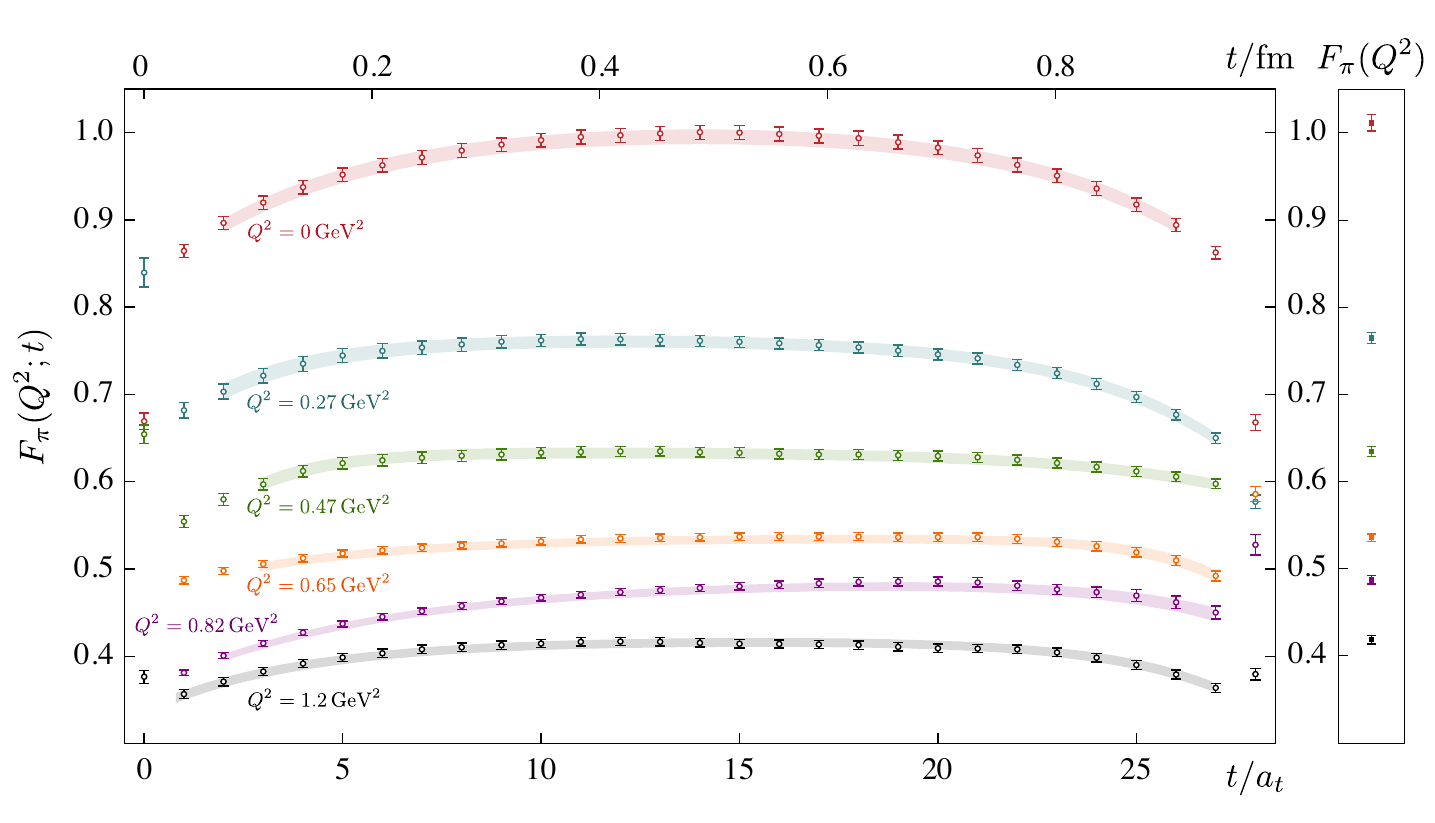}
  \caption{ Typical $F_\pi(Q^2; t)$ extracted from optimized three-point functions (points) with fit descriptions using Eqn.~\ref{three_point_fit} (curves). Note that the data points have a high degree of timeslice correlation which is accounted for in the fitting.  \label{fig::pion_proj0_pion_proj0_Q2_dep}}
\end{figure}

The pion form-factor appears in the matrix element decomposition, $\big\langle \pi^+(\vec{p}\,') \big| j^\mu \big| \pi^+(\vec{p}) \big\rangle = (p+p')^\mu \, F_\pi(Q^2)$, which we will extract from three-point Euclidean correlation functions computed using optimized ground-state pion operators of definite momentum at the source (at $t=0$) and the sink (at $\Delta t = 28\, a_t$). As discussed previously, we will present $F(Q^2;t)$, where the leading Euclidean time-dependence of the correlation function has been removed, with any remaining time-dependence signaling the presence of excited state contributions to the correlation function. By utilizing many values of $\vec{p}$ and  $\vec{p}\,'$ we can determine the form-factor at a range of $Q^2$ values. We plot $F_\pi(Q^2;t)$ for a subset of these $Q^2$ values in Figure~\ref{fig::pion_proj0_pion_proj0_Q2_dep}, where for each $Q^2$ we overlay a fit according to the form in Eq.~\ref{three_point_fit}.

\begin{figure*}
  \includegraphics[width=0.84\linewidth]{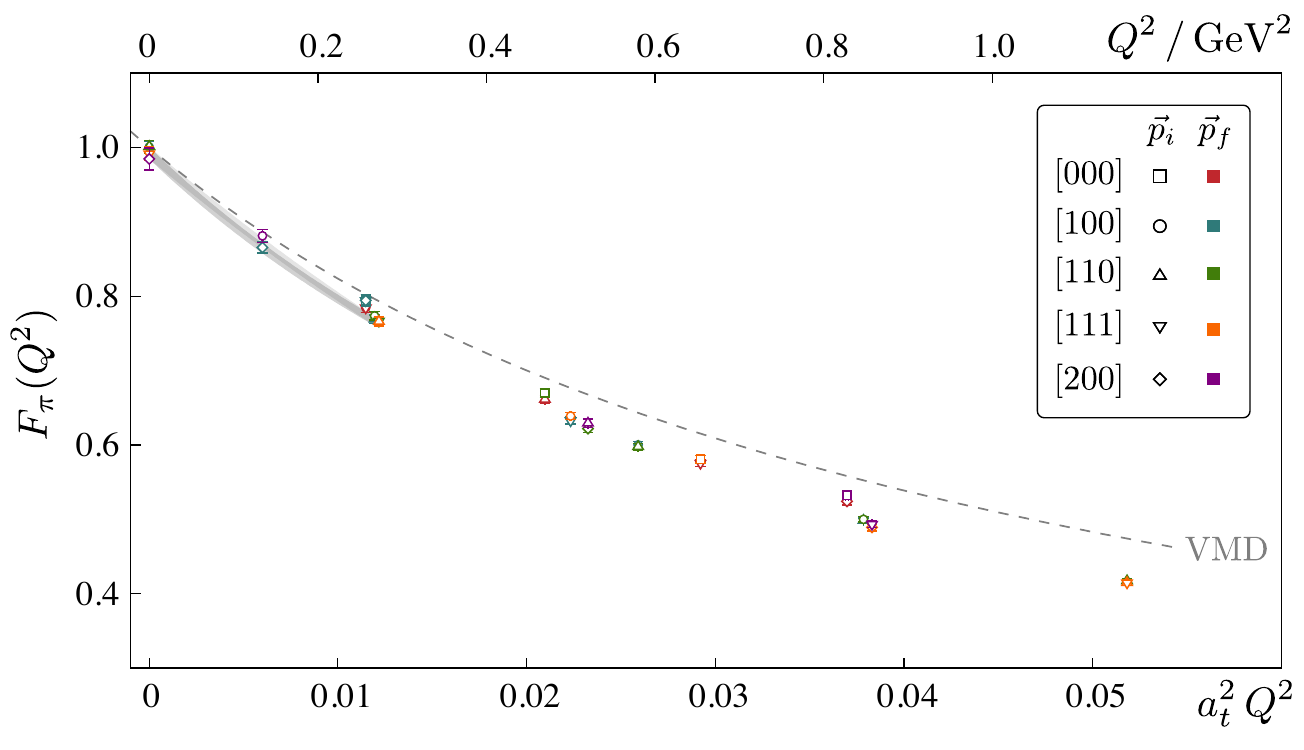}
  \caption{ Pion ground-state form-factor, $F_\pi(Q^2)$, using the improved current, Eqn.~\ref{improvement}. Vector meson dominance using the $\rho$ meson mass on this lattice shown by the dashed curve. Fits to the small-$Q^2$ points using gaussian and single-pole forms shown by the gray curves.
  \label{fig::pion_formfactor} }
\end{figure*}

\begin{figure*}
  \includegraphics[width=0.90\linewidth]{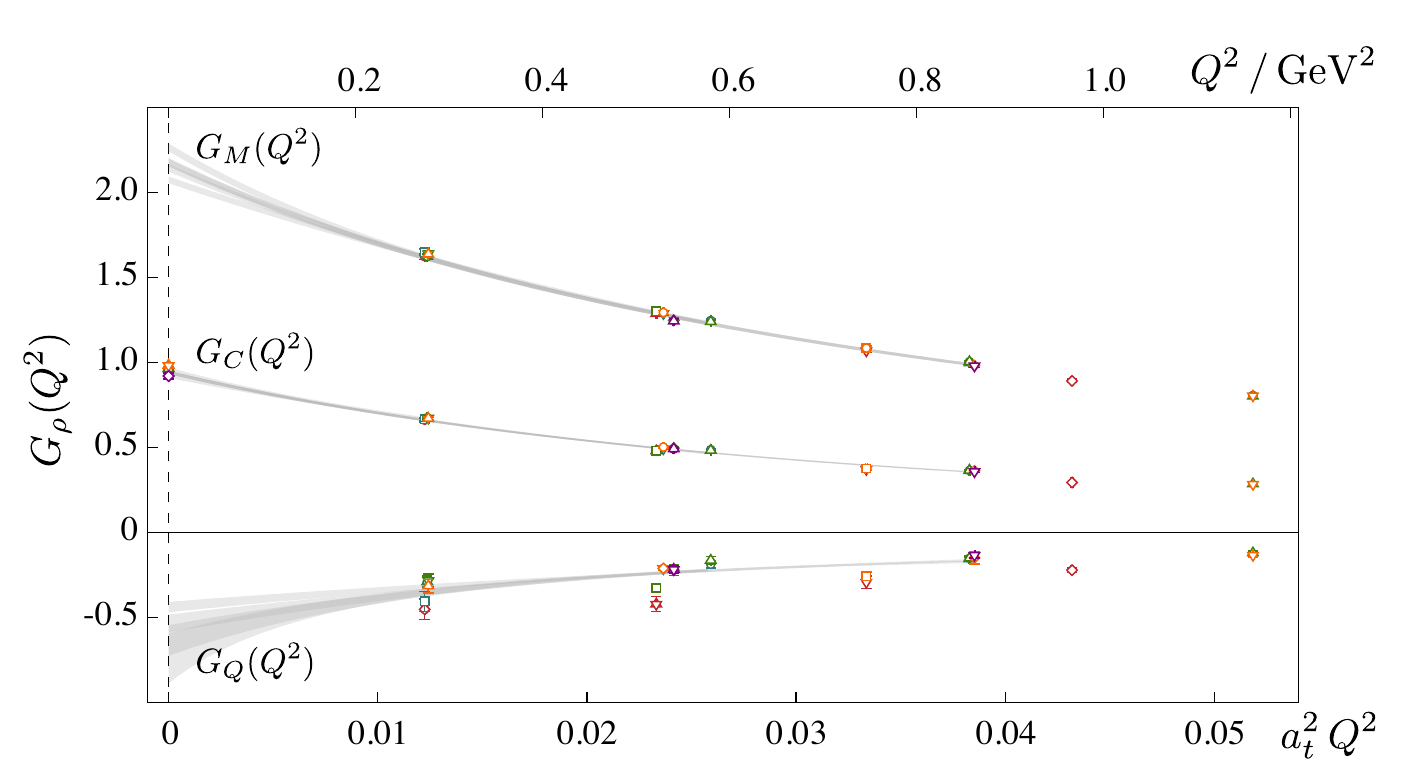}
  \caption{ Ground-state $\rho$ meson multipole form-factors. Points have the same color and shape labeling presented in Figure~\ref{fig::pion_formfactor}. Fits to the $Q^2$ dependence, described in the text, are shown as gray curves. 
 \label{fig::rho_form_factors}}
\end{figure*}

In Figure~\ref{fig::pion_formfactor} we plot the resulting $Q^2$ dependence, shown via both dimensionless $a_t^2 Q^2$ and scale-set using the $\Omega$-baryon mass prescription presented in Section~\ref{sec::calc_details}. A large number of kinematic points are sampled by considering all combinations of momentum such that $n^2_{\vec{p}} \le 4$, $ n^2_{\vec{p}\,'} \le 4$ and $ n^2_{\vec{q}} \le 4$. The extracted points, for the improved current discussed in Section~\ref{ssec::three_points_improv}, appear to lie on a single curve, with only small residual scatter which can originate from fitting-range systematics and modest discretization effects.

Describing the $Q^2$ dependence may offer some phenomenological insight, albeit in this calculation at an unphysically heavy quark mass. A commonly used approach to describe vector-current form-factors of hadrons is to argue that the photon is behaving like the lightest vector meson which can couple to the hadrons in question, which in this case would be the $\rho$. This ``vector meson dominance"(VMD) describes the $Q^2$ dependence by ${F_{\mathrm{VMD}}(Q^2) = \frac{1}{1 + Q^2/m_\rho^2 }}$. Using the $\rho$ mass determined on these lattices, $m_\rho = 1020(1)\, \mathrm{MeV}$, we have the dashed curve shown in Figure~\ref{fig::pion_formfactor}, which is seen to describe the lattice data reasonably well only for small photon virtualities. One possible explanation of this effect is that as we move out to larger $Q^2$, considering only the nearest time-like pole, the $\rho$, and neglecting all excitations, becomes a progressively poorer approximation.

The distribution of charge within the pion can be characterized by the \emph{charge radius}, defined via the slope of the form-factor at zero virtuality, ${\langle r^2 \rangle \equiv -6 \frac{d}{dQ^2}F(Q^2) \big|_{Q^2=0}}$. We may obtain this quantity from the discrete $Q^2$ data presented in Figure~\ref{fig::pion_formfactor} by parameterizing the $Q^2$--dependence for small virtualities. Considering gaussian ${ \big(  F_\pi(Q^2) = F(0)\, e^{-Q^2/16\beta^2}  \big) }$ and pole ${ \big(  F_\pi(Q^2) = F(0)\, \tfrac{1}{1 + Q^2/m^2} \big) }$ forms to describe ${Q^2 < 0.3\,\mathrm{GeV}^2}$, we obtain\footnote{If $F(0)$ is allowed to float in fits, a value statistically compatible with 1 is obtained, as it must since the pion form-factor at zero $Q^2$ was used to set $Z_V$. The fit $\chi^2$ values obtained are fairly large due to the scatter in the statistically precise data, which is likely due to small discretization effects which are not described by these smooth fit-forms.} 
a charge radius ${  \langle r^2 \rangle_{\pi}^{1/2} = 0.47(6) \; \mathrm{fm}  }$, where the error includes the variation over fit-form. As we might expect, in a calculation where three flavors of quarks all have approximately the strange quark mass, we obtain a pion charge radius somewhat smaller than the physical pion ${\langle r^2 \rangle_\pi^{1/2} = 0.67(1)\,\mathrm{fm}}$ \cite{Amendolia:1986wj, PDG-2012}, and also smaller than the physical kaon ${  \langle r^2 \rangle_K^{1/2} = 0.58(4) \, \mathrm{fm}  }$ \cite{Amendolia:1986ui}.

\subsubsection{$\rho$ form-factors}

The three form-factors required to describe the vector-current response of a vector hadron may be defined as in Eq.~\ref{eqn::rho_Gmultipole_basis}, which makes use of a multipole basis. The decomposition presented in Eq.~\ref{eqn::rho_Gi_basis} defines the linear system which we may solve, as described in Section~\ref{sec::3pt_lin_sys}, for the form-factors. We plot the charge, $G_E(Q^2)$, magnetic, $G_M(Q^2)$, and quadrupole, $G_Q(Q^2)$ form-factors in Figure~\ref{fig::rho_form_factors}. Examination of Eqs~\ref{eqn::rho_Gi_basis}, \ref{eqn::rho_Gmultipole_basis} indicates that only the charge form-factor has a non-zero kinematic factor when $Q^2=0$, and as such only it is determined there, while all three form-factors are sampled for positive non-zero $Q^2$. The smallest form-factor, $G_Q$, shows the largest scatter, which likely originates from modest discretization effects and timeslice fitting-range fluctuations.

Fitting the $Q^2$ dependence of the charge form-factor with various forms\footnote{$G(0) e^{-Q^2/16\beta^2}$, $G(0) e^{-Q^2(1+\alpha Q^2)/16\beta^2}$, $G(0)/(1+ Q^2/m^2)$, $G(0)/(1+ Q^2/m^2 + \gamma (Q^2/m^2)^2 )$, $G(0) \frac{e^{-Q^2/16\beta^2}  }{1+Q^2/m^2}$ }, over various $Q^2$ ranges we obtain $G_C(0) = 0.94(1)$ and ${  \langle r^2 \rangle_\rho^{1/2} = 0.55(5) \, \mathrm{fm} }$ where the errors include a systematic variation over different fit forms. The deviation of the charge from 1 was discussed previously in Section~\ref{ssec::three_points_improv}. 

In order to determine the magnetic and quadrupole moments from $G_M(0)$ and $G_Q(0)$ it is necessary to parameterize the $Q^2$ dependence of the form-factors and extrapolate back to $Q^2=0$. Utilizing a range of possible forms, we obtain ${G_M(0) = 2.17(10)}$ and ${G_Q(0) = -0.54(10)}$, accounting for the variation over fit-forms, which is much larger than the statistical uncertainty, in the errors. More precise determinations of these quantities could be obtained if twisted boundary conditions were used to sample the form-factors at smaller $Q^2$ (see for example \cite{Flynn:2005in}).

Within a simple picture of the $\rho$ as a $q\bar{q}$ bound-state, the presence of a quadrupole moment would indicate a required admixture of $D$-wave into the dominantly $S$-wave wavefunction. Previous estimates of the $\rho$-meson magnetic moment in versions of QCD with heavier than physical quarks come from chiral effective theory \cite{Djukanovic:2013mka} where $G_M(0) \sim 2.2$ for large pion masses, and quenched lattice QCD using either an energy shift in a magnetic field \cite{Lee:2008qf} where $G_M(0) = 2.13(6)$, or extrapolation to zero $Q^2$ from a single spacelike virtuality \cite{Hedditch:2007ex} where $G_M(0) = 2.05(4)$, at comparable unphysical pion masses. A dynamical calculation, Ref.~\cite{Owen:2015gva}, which appeared while this manuscript was in the final stages of production, found, at a comparable pion mass, $G_M(0) = 2.23(2)$ and $G_Q(0) = -0.362(20)$, using a model extrapolation to $Q^2=0$ from a single non-zero $Q^2$ point.

\subsubsection{$\pi'$ form-factor}

\begin{figure}[b]
  \includegraphics[width=\linewidth]{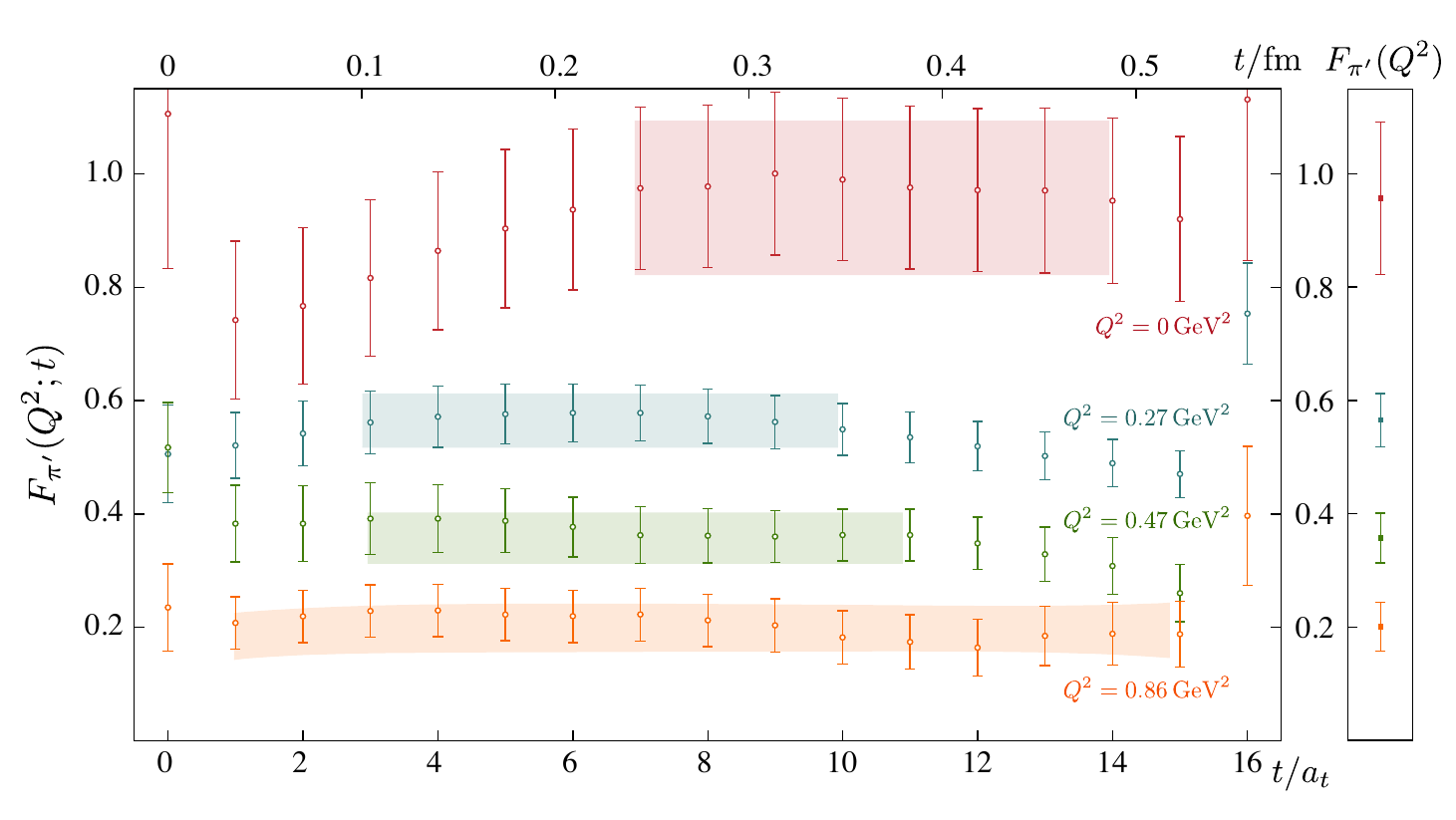}
  \caption{Current insertion time dependence for the form-factor of the first excitation of the pion, $F_{\pi'}(Q^2;t)$, shown for a range of source and sink momenta. The high degree of timeslice-timeslice data correlation  is manifested in the error on the fit which is not significantly reduced relative to the error on the individual data points. \label{fig::pion_proj1_pion_proj1_Q2_dep}}
\end{figure}

The examples presented in the previous two subsections were the lightest states with the relevant quantum numbers. As such it was not strictly necessary to use optimized operators -- any suitable meson interpolators used in the three-point functions will, in the limit of large time separations, give access to the matrix elements. We will now move to the case of an excited state, the first excitation of the pion, which we access using optimized operators to eliminate the contribution of the ground-state pion.

As described in Section \ref{sec::three-point}, the signals for excited states are typically noisier than those for the ground state, and as such we separate the source and sink operators by a smaller time, in this case $\Delta t = 16 \, a_t$. The decomposition for this matrix element is of the same form as the pion described previously, Eq.~\ref{pion_ff}. We plot the extracted form-factor, $F_{\pi'}(Q^2; t)$, as a function of the current insertion timeslice in Figure~\ref{fig::pion_proj1_pion_proj1_Q2_dep}.

\begin{figure}[h]
  \includegraphics[width=1.07\linewidth]{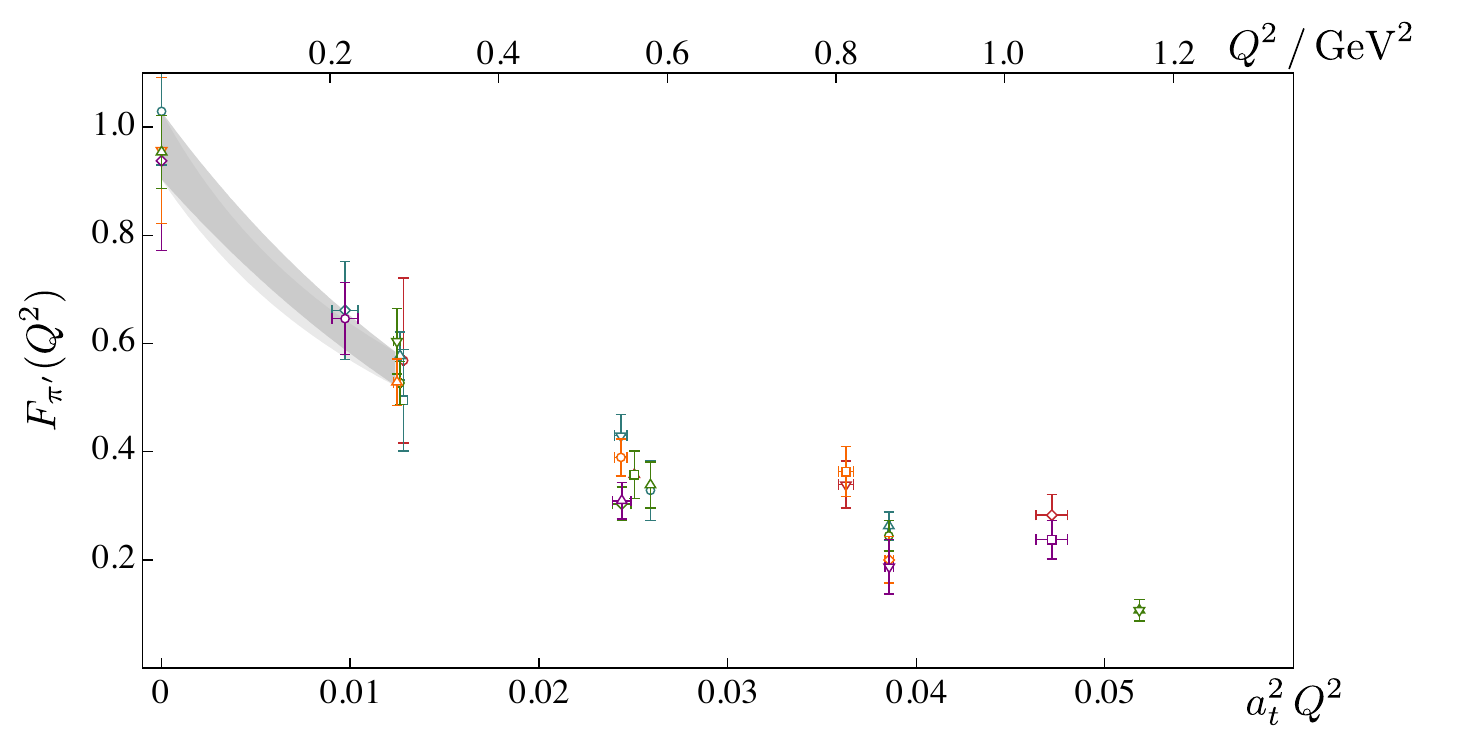}
  \caption{First-excited pion form-factor, $F_{\pi'}(Q^2)$. Points have the same color and shape labeling presented in Figure~\ref{fig::pion_formfactor}. Fits to low $Q^2$ dependence used to constrain charge-radius, as described in text, shown as gray bands.  
  \label{fig::piStar_form_factor}}
\end{figure}

The $Q^2$ dependence of the form-factor, $F_{\pi'}(Q^2)$, is presented in Figure~\ref{fig::piStar_form_factor}. While the extracted values at $Q^2=0$ are not statistically precise, they are certainly consistent with unity. The charge radius can be extracted from the slope at $Q^2=0$ which we determine by parameterizing\footnote{
Gaussian ($F_{\pi'}(0) \, e^{-Q^2/16\beta^2}$) and one-pole ($F_{\pi'}(0) / (1 + Q^2 / m^2 )$) forms were used.  
}
 the data for $Q^2 \lesssim 0.3\,\mathrm{GeV}^2$, yielding ${   \langle r^2 \rangle_{\pi'}^{1/2} = 0.74(6) \, \mathrm{fm} }$ where the error includes variation over parameterization form. As we might expect for a state which likely can be characterized as a radial excitation, this is significantly larger than the $0.47(6)\,\mathrm{fm}$ found for the ground-state pion at this quark mass. 
 
Ref.~\cite{Owen:2015gva}, computing at a very similar pion mass found $0.517(4)\,{\rm fm}$ for the ground-state pion charge radius, and $0.59(3)\,{\rm fm}$ for the first excitation of the pion. Their approach determines a single point on the form-factor curve at $Q^2\sim 0.16 \,\mathrm{GeV^2}$ which is used to determine the slope at $Q^2=0$ assuming monopole dependence on $Q^2$.

\subsection{Radiative Transitions}

\subsubsection{$\pi' \rightarrow \pi \gamma$ transition}

\begin{figure}[h]
  \includegraphics[width=1.05\linewidth]{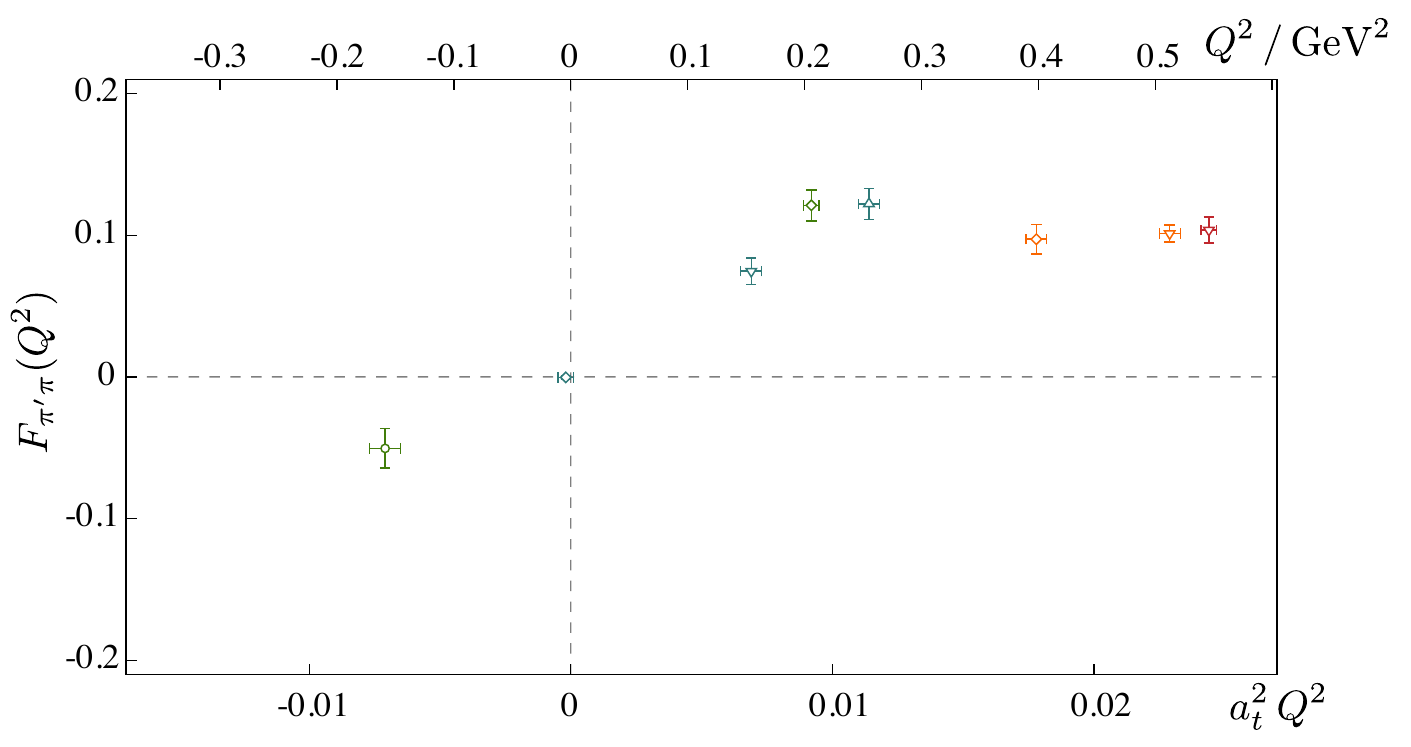}
  \caption{Transition form-factor between first-excited and ground-state pions, $\pi' \to \pi \gamma$. Points have the same color and shape labeling presented in Figure~\ref{fig::pion_formfactor} with the excited pion having momentum $\vec{p}_f$. 
  \label{fig::pi_pistar_transition}}
\end{figure}

In a transition between different pseudoscalar mesons, the decomposition of the current in terms of a form-factor $F_{\pi' \pi}(Q^2)$ is as in Eqn.~\ref{pi_pi_ff}, and the form-factor must vanish at $Q^2=0$. The transition form-factor is extracted from three-point functions with $\Delta t = 20 \, a_t$, fitting the time-dependence as previously to account for any residual unwanted excited state contribution. We plot the extracted form-factor in Figure~\ref{fig::pi_pistar_transition} -- that we are now able to explore the timelike $Q^2$ region, where previously all points were spacelike, follows from the differing masses of the hadrons at source and sink, a simple example being the case where $\vec{p}\,' = \vec{p}$, so that ${Q^2 = - \big( E'(\vec{p}\,) - E(\vec{p}\,) \big)^2 < 0}$. In order to be able to trivially relate our Euclidean amplitudes to Minkowski amplitudes, we must restrict ourselves to the region where the current is not timelike enough to produce on-shell hadrons. In this calculation where the $\pi \pi$ threshold is above the $\rho$ mass, this limits us to ${  Q^2 > - m_\rho^2 \sim - 1\,\mathrm{GeV}^2  }$. In order to explore further into the timelike region, a somewhat more sophisticated approach must be followed \cite{Meyer:2011um, Feng:2014gba}.

\subsubsection{$\rho \rightarrow \pi \gamma$ transition}\label{rhopigamma}

\begin{figure*}
  \includegraphics[width=0.71\linewidth]{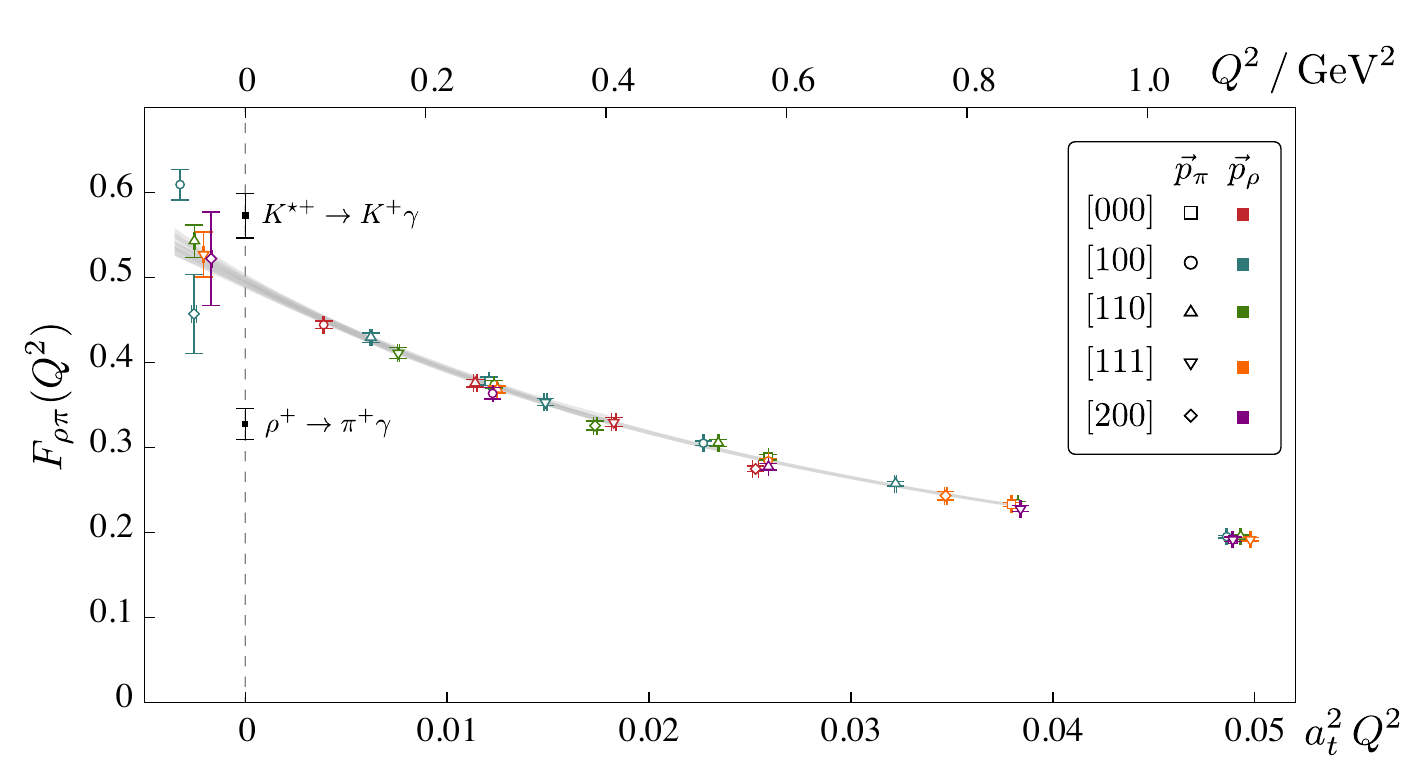}
  \caption{Ground-state $\rho$ to ground-state $\pi$ transition form-factor. Curves in gray show fits used to interpolate between spacelike and timelike regions to determine the photocoupling, $F_{\rho\pi}(0)$. Experimental decay widths converted to photocouplings shown for orientation.  \label{fig::rho_pi_transition}}
\end{figure*}

A suitable decomposition for a vector to pseudoscalar transition in terms of a dimensionless form-factor is given in Eqn.~\ref{rho_pi_ff}. Using optimized operators for the ground-state $\rho$ and ground-state $\pi$ we computed correlation functions with $\Delta t = 28\, a_t$ for a large range of source and sink momenta -- the resulting determination of the form-factor, $F_{\rho \pi}(Q^2)$ is presented in Figure~\ref{fig::rho_pi_transition}.

The value of the form-factor at $Q^2=0$, known as the photocoupling, is of particular interest since it controls the rate of the physically allowed radiative transition process, $\rho^\pm \to \pi^\pm \gamma$. As can be seen in Figure~\ref{fig::rho_pi_transition}, we do not determine this quantity directly, but we may estimate it using interpolation between our space-like and time-like points. Using a range of fit forms over several $Q^2$ ranges (plotted in grey) we estimate $F_{\rho \pi}(0) = 0.494(8)$, where the error includes variation over fit-forms.

The Lorentz invariant matrix element for the decay $\rho^+ \to \pi^+ \gamma$ can be obtained by contracting the matrix element in Eqn.~\ref{rho_pi_ff} with a final state polarization vector, $\mathcal{M}_{\lambda_\gamma,\lambda} = \epsilon^*_\mu(\lambda_\gamma,\vec{q}) \big\langle \pi^+(\vec{p}\,') \big| j^\mu \big| \rho^+(\lambda, \vec{p}) \big\rangle$,
and for a vector stable under the strong interaction, we may obtain the decay width from
\begin{equation*}
\Gamma(\rho^+\rightarrow \pi^+\gamma) = \frac{1}{32\pi^2}\int \!\! d\Omega_{\vec{q}} \, \frac{\lvert \vec{q}\rvert}{m_\rho^2} \,  \frac{1}{3} \sum_{\lambda_\gamma,\lambda} \lvert \mathcal{M}_{\lambda_\gamma,\lambda}  \rvert^2,
\end{equation*}
where we have summed over the final state photon polarizations and averaged over the initial state polarization of the $\rho$. Using the decomposition above, and restoring the factors of $e$, we obtain the result relating the width to the photocoupling,
\begin{equation*}
\Gamma(\rho^+\rightarrow \pi^+\gamma) = \alpha\frac{4}{3} \frac{\lvert \vec{q} \rvert^3}{(m_\rho \!+\! m_\pi)^2} \lvert F_{\rho \pi}(0) \rvert^2,
\end{equation*}
where $\alpha = e^2/4\pi$. 

The calculation performed here uses three degenerate quark flavors tuned to approximate the physical strange quark mass and as such our photocoupling determination cannot be directly compared with experiment. For orientation we show in Figure~\ref{fig::rho_pi_transition}, the experimental values, $F_{\rho \pi}(0) = 0.33(2)$ and $F_{K^{\!*}\!K}(0) = 0.57(3)$ extracted from the corresponding decay rates obtained via the Primakoff effect for pions and kaons incident on nuclear targets \cite{PhysRevD.33.3199,Capraro:1987rp,PhysRevLett.51.168}.

The $Q^2$ dependence of this meson transition form-factor plays a role in models of deuteron electromagnetic structure, where a virtual photon probe may couple to the bound nucleons or to the meson currents proposed to supply the binding \cite{Arnold:1979cg}.

\subsubsection{$\rho'\rightarrow\pi\gamma$ transition}

The first-excited $\rho$ state may also undergo a transition to the ground-state pion, with the form of the decomposition of the matrix element being the same as in the previous section. In Section~\ref{sec::two_points} we presented the spectrum of excited vector mesons, finding that the first-excited state, $m_{\rho'} = 1882(11) \,\mathrm{MeV}$, is close to being degenerate with the second-excited state $m_{\rho''} = 1992(6) \,\mathrm{MeV}$. Our use of optimized operators corresponding to orthogonal combinations of basis operators allows us to reliably study the two excitations independently. 

\begin{figure}[h]
  \includegraphics[width=\linewidth]{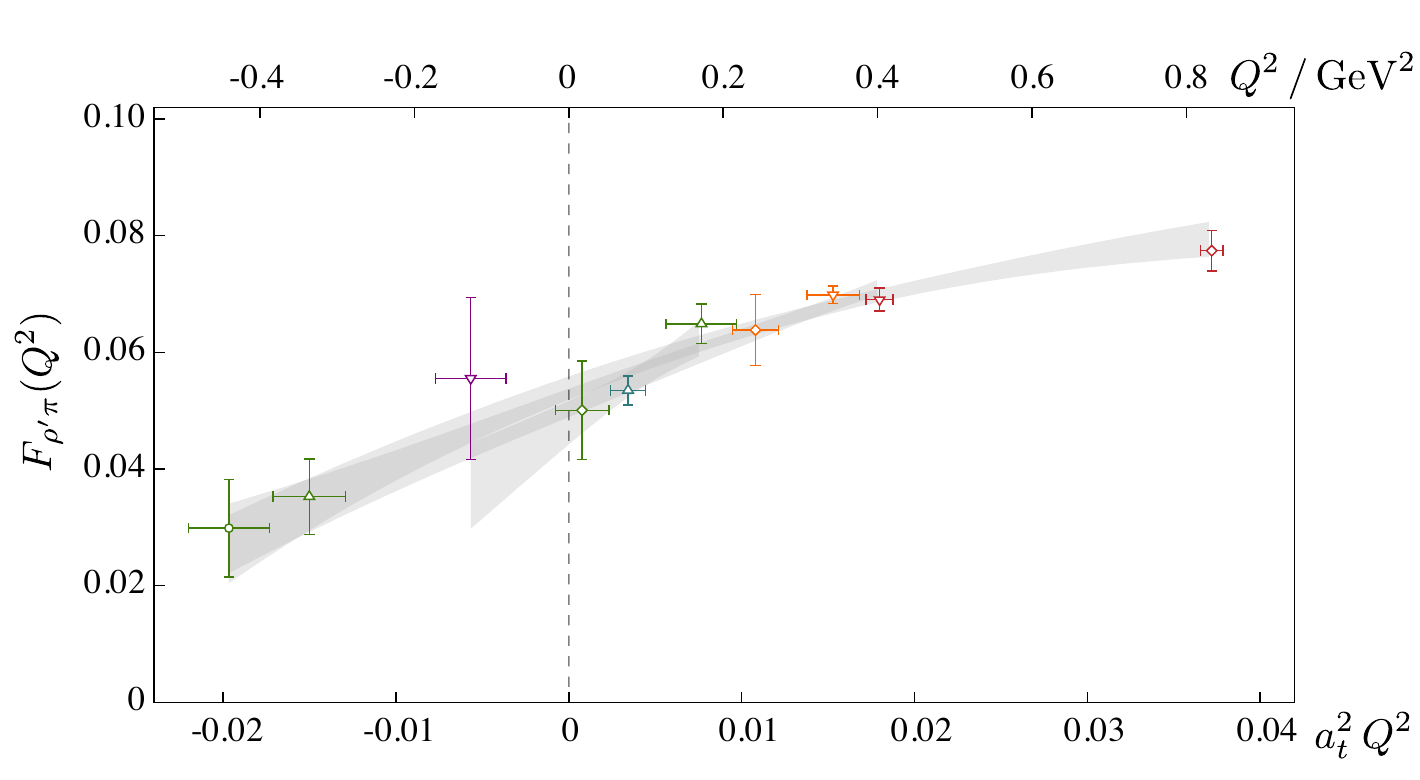}
  \caption{First-excited $\rho$ transition to ground-state $\pi$ form-factor, $F_{\rho'\pi}(Q^2)$. Points have the same color and shape labeling presented in Figure~\ref{fig::rho_pi_transition}. Gray curves show fits used to interpolate to the photocoupling. \label{fig::rho1_pi_transition}}
\end{figure}

We extract the form-factor using optimized operators in correlation functions with time-separation, $\Delta t = 20 \,a_t$, with the results presented in Figure~\ref{fig::rho1_pi_transition}. To determine the photocoupling, $F_{\rho' \pi}(0) = 0.050(4)$, we perform fits to the data over various $Q^2$ ranges using several fit-forms, and the quoted uncertainty includes this variation.

The photocoupling for this transition is observed to be an order of magnitude smaller than that of $\rho \to \pi \gamma$ extracted in Section~\ref{rhopigamma}. Within simple models treating mesons as $q\bar{q}$ bound-states with non-relativistic wavefunctions, such a suppression is expected -- the net effect of the current is to slightly shift in momentum-space the wavefunction of the pion, and since the $\rho'$ is likely described as a radial excitation, the resulting wavefunction overlap is much reduced relative to that for the ground-state $\rho$. This is described as a `hindered' magnetic dipole transition. A relevant experimental example of a hindered transition lies in the charmonium sector -- the relative rates of $\psi(2S) \to \eta_c \gamma$ and $J/\psi \to \eta_c \gamma$, $\frac{\Gamma(\psi(2S) \to \eta_c \gamma)/|\vec{q}_{\psi(2S)}|}{\Gamma(J/\psi \to \eta_c \gamma)/|\vec{q}_{J/\psi}|} \sim 0.1$ show the expected hierarchy of hindered versus non-hindered~\cite{PDG-2012}.

\subsubsection{$\rho''\rightarrow \pi \gamma$ transition}

\begin{figure}[h]
  \includegraphics[width=\linewidth]{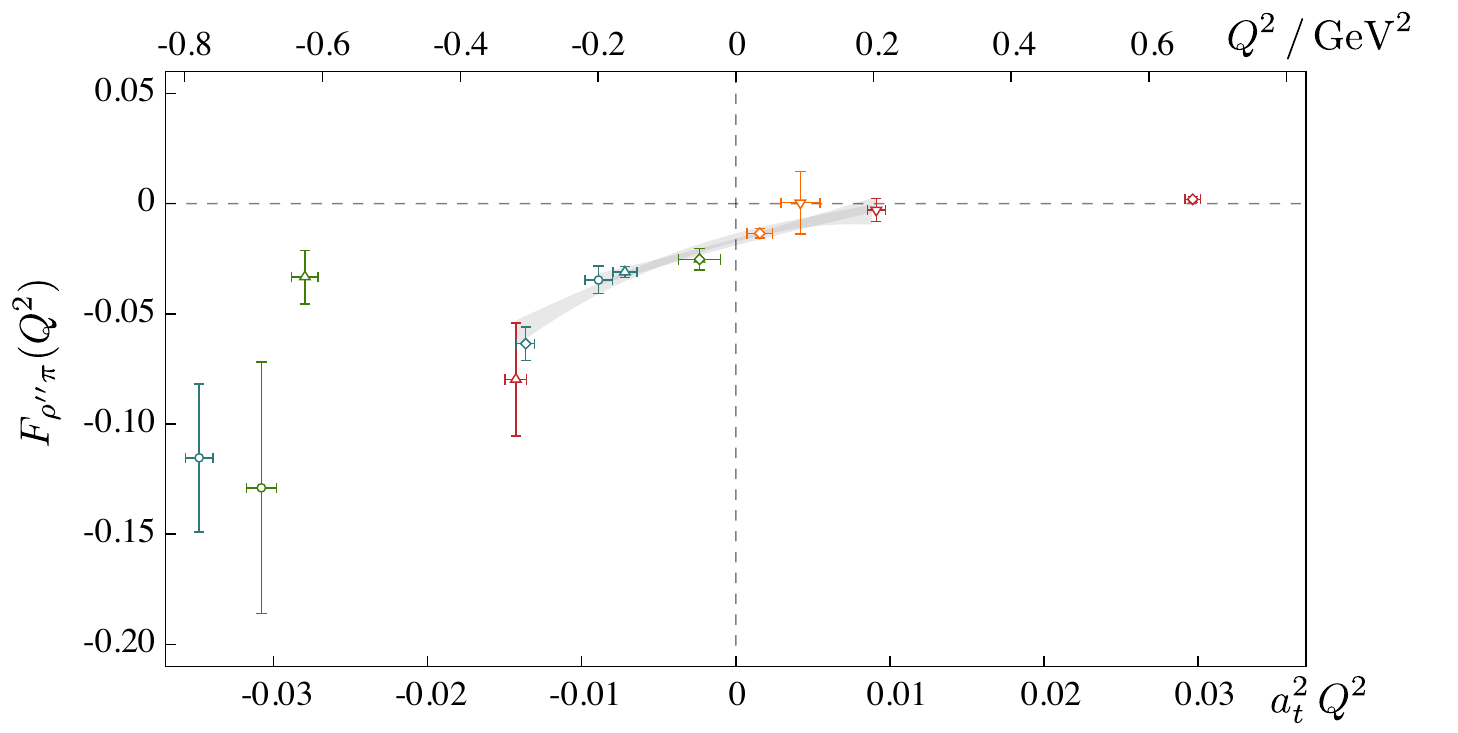}
    \caption{ Second-excited $\rho$ transition to ground-state $\pi$ form-factor, $F_{\rho'' \pi}(Q^2)$. Points have the same color and shape labeling presented in Figure~\ref{fig::rho_pi_transition}. Gray curves show fits used to interpolate to the photocoupling. 
     \label{fig::rho2_pi_transition}}
\end{figure}

An extraction analogous to that presented in the previous subsection can be performed for the second-excited $\rho$ state, leading to the form-factor shown in Figure~\ref{fig::rho2_pi_transition}. Interpolating to $Q^2=0$ using a range of forms yields $F_{\rho'' \pi}(Q^2) = -0.016(3)$, which is smaller still than the $\rho' \to \pi \gamma$ photocoupling. The sign is somewhat arbitrary and would only have definite meaning were we to compare to other transitions involving the $\rho''$.

Within a simple $q\bar{q}$ bound-state model we might expect the $\rho''$ state to be dominated by a $^3\!D_1$ configuration (and indeed the operator overlaps presented in \cite{Dudek:2011bn} seem to suggest this), which would have a `hindered' structure in a transition to the ground-state $S$-wave pseudoscalar owing to the need for the current to provide a $D$-wave angular dependence, which appears only as a relativistic correction to the leading behavior. 

\subsubsection{$\pi' \rightarrow \rho \gamma$ transition}

\begin{figure}[h]
  \vspace{-0.5cm}
  \includegraphics[width=\linewidth]{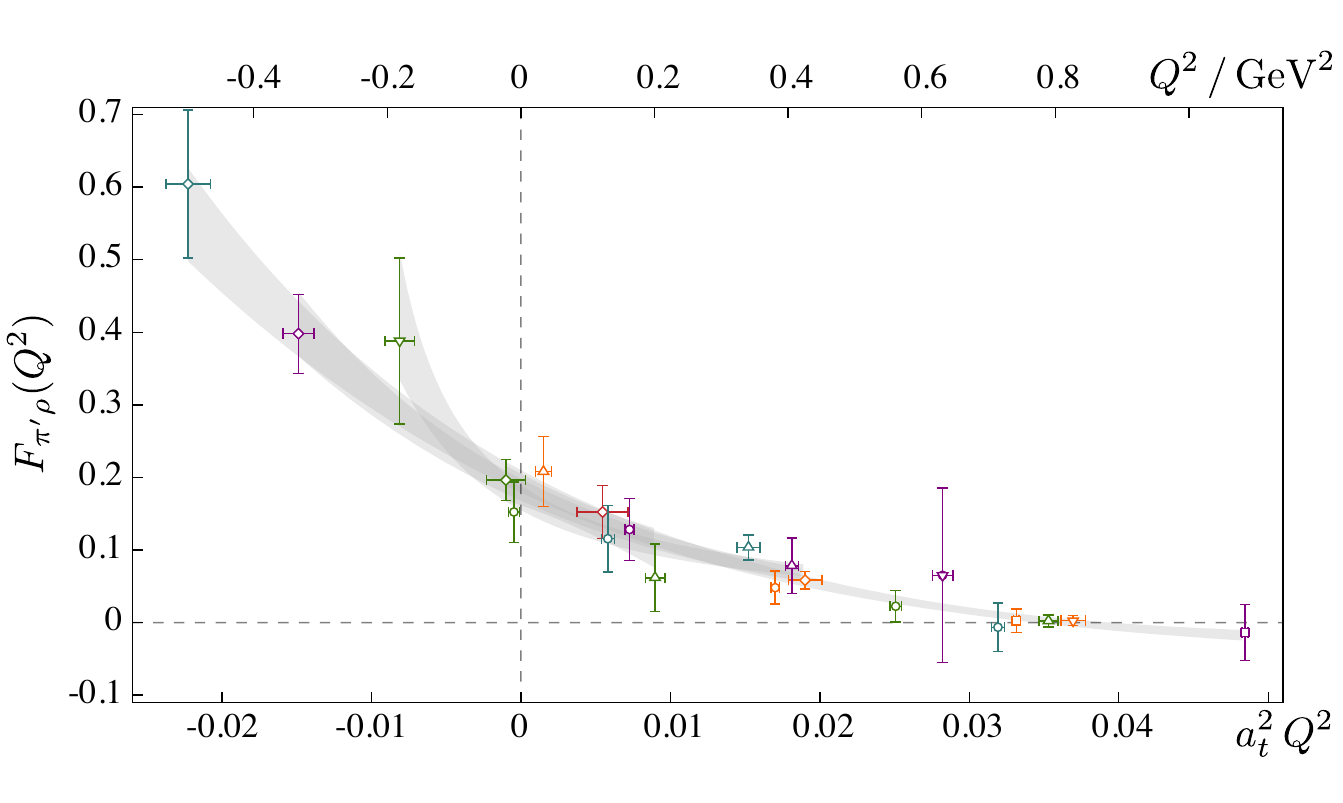}
  \vspace{-0.5cm}
    \caption{First-excited $\pi$ transition to ground-state $\rho$, $F_{\pi' \rho}(Q^2)$. Gray curves show fits used to interpolate to the photocoupling. \label{fig::pi1_rho0_transition}}
\end{figure}

The first-excited pion may undergo a transition to the ground-state $\rho$. The results, extracted from $\Delta t = 20 \, a_t$ correlation functions, are presented in Figure~\ref{fig::pi1_rho0_transition}, along with a number of parameterizations used to interpolate a photocoupling of $F_{\pi' \rho}(0) = 0.18(2)$. Again we observe a significant suppression relative to the $\rho \to \pi \gamma$ case in line with this being a hindered transition.

\subsubsection{$\rho' \rightarrow \pi' \gamma$ transition}

\begin{figure*}
  \includegraphics[width=0.7 \linewidth]{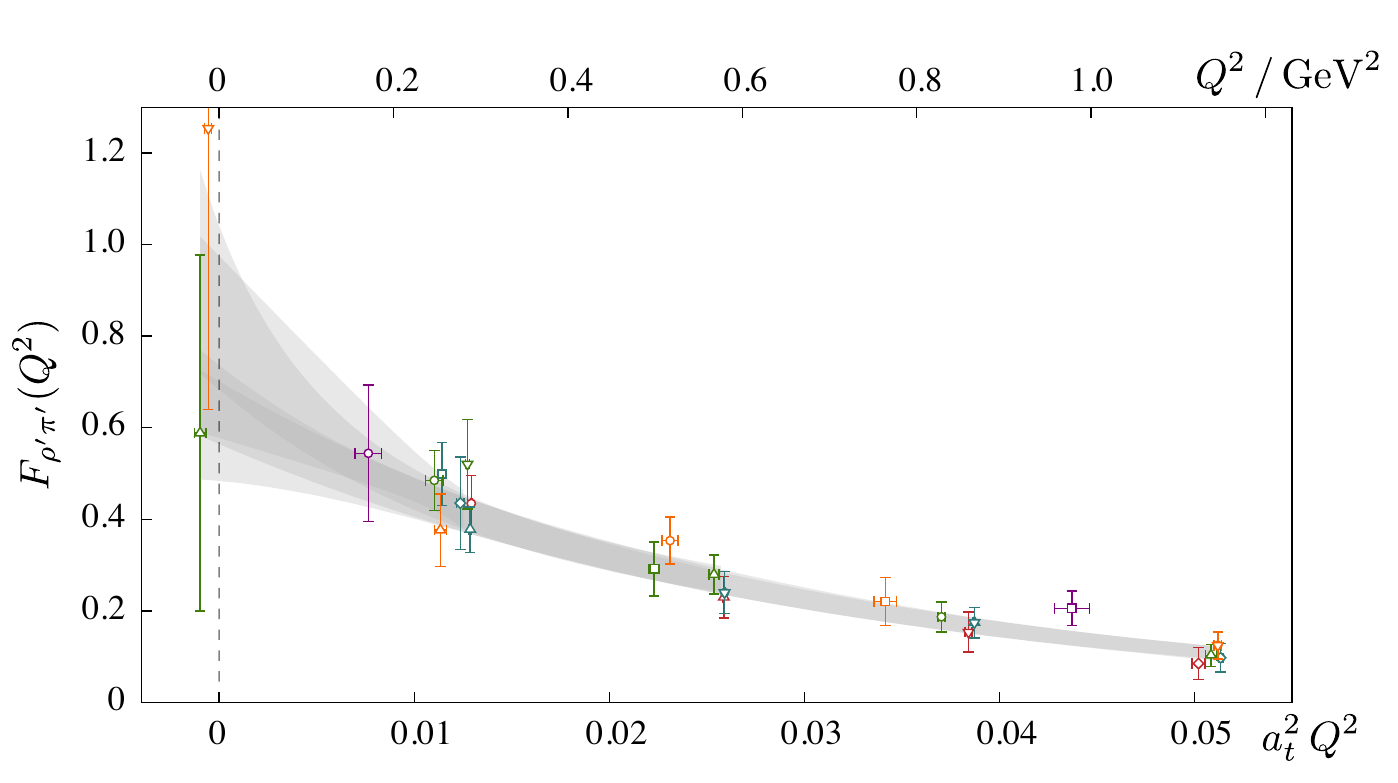}
      \caption{First-excited $\pi$ transition to first-excited $\rho$, $F_{\pi' \rho'}(Q^2)$. Gray curves show fits used to interpolate to the photocoupling. \label{fig::rho1_pi1_transition}}
\end{figure*}

This transition, which occurs between excited states, is not expected to be hindered in the case that the $\rho'$ and $\pi'$ are identified predominantly as the first radial excitations of the $\rho$ and $\pi$ respectively. As such we might expect a somewhat larger photocoupling than in previous subsections. We extracted the form-factor from optimized operator correlation functions with $\Delta t = 20 \,a_t$ obtaining the results presented in Figure~\ref{fig::rho1_pi1_transition}. Fits to the $Q^2$ dependence with a range of forms lead to an estimate of the photocoupling, $F_{\rho' \pi'}(Q^2) = 0.7(2)$, which, although not determined with high precision, is of comparable size to the $\rho \to \pi \gamma$ coupling. 

\section{Summary \label{sec::summary}}

A desire to extract current matrix elements between excited hadrons motivated our exploration of \emph{optimized operators} which are capable of interpolating only a single hadron eigenstate from the vacuum rather than a superposition of all possible eigenstates.

We have demonstrated through explicit calculation, the utility of these optimized operators when used at the source and sink of three-point correlation functions also featuring a vector current insertion. Matrix elements featuring the two lightest pseudoscalar mesons and the three lightest vector mesons were explored, with successful extraction of excited-state transition matrix elements at a range of momentum transfers. In the case that the optimized operators correspond to ground-states, their use reduces the degree of unwanted excited-state contribution to the correlation function and allows for the source and sink to be separated by a smaller time interval with a corresponding reduction in statistical noise.

Optimized operators, as we have implemented them, are linear superpositions of a large basis of meson operators, and it follows that we need a technique that allows efficient evaluation of potentially complicated operator constructions. \emph{Distillation} has previously been shown to meet these needs in the case of two-point correlation functions, and in this study we have demonstrated its efficacy in the case of three-point functions, the first time it has been used in this manner. Its use allows the problem to be factorized into pieces corresponding to operator constructions of definite momentum at source and sink and independent pieces corresponding to quark propagation. This separation of operators from propagators avoids the problem encountered in sequential-source techniques of requiring the operators to be selected before the Dirac matrix inversions take place.

This first study was restricted to transitions involving pseudoscalar and vector mesons, but we may easily extend it to other meson quantum numbers using the flexible operator basis presented in \cite{Dudek:2010wm, Dudek:2009qf, Thomas:2011rh}. One important application is the determination of transition photocouplings relevant to meson photoproduction experiments like GlueX and CLAS12, where the process is modeled as proceeding through $t$-channel meson exchange. Part of the role of these experiments is to search for exotic and non-exotic hybrid mesons, those states which have an essential gluonic contribution in their wavefunction -- it has been claimed previously in models \cite{Isgur:1985vy, Isgur:1999kx, *Close:2003fz, *Close:2003ae} that transition photocouplings of hybrid mesons may be large, motivating photoproduction searches. The technology we have explored for mesons may also be applied to the baryon sector in which the dependence of the transition form-factors on photon virtuality can be measured in electroproduction experiments on proton and neutron targets \cite{Aznauryan:2012ba}.

Throughout this analysis we have proceeded under the assumption that the hadrons we consider are stable eigenstates of QCD. In fact, for the quark mass used in the calculation only the ground-state $\pi$ and $\rho$ are below all relevant kinematic thresholds, while the various excitations can in principle decay. As the light quark masses are reduced toward their physical value, even the ground state $\rho$ ceases to be an eigenstate and rather appears as a \emph{resonance} in $\pi\pi$ scattering. In recent years there has been significant progress determining meson resonance properties using lattice QCD, making use of the  discrete spectrum of states in the finite volume defined by the periodic lattice boundary, following the formalism initially presented in \cite{Luscher:1990ux}. 

To date, however, there is no calculation exploring the coupling of a resonance to external currents. A rigorous calculation at physical kinematics, where the $\rho$ is a resonance, seeking the coupling $\rho \to \pi \gamma$ would in fact need to determine the $P$-wave partial-wave amplitude for $\pi \gamma \to \pi \pi$ as a function of the invariant mass, $m_{\pi\pi}$. By analytically continuing the amplitude to complex values of $m^2_{\pi\pi}$ and extrapolating to the $\rho$-resonance pole, the coupling could be extracted as the residue of the amplitude. Only very recently \cite{Briceno:2014uqa} has the formalism relating matrix elements extracted in a finite-volume to the physical amplitude been presented.

The techniques laid out in this paper, which allow the extraction of matrix-elements for each state in a tower of discrete eigenstates, will be required in any attempt to determine resonance couplings to external currents. The extension of the operator basis to include, as well as single-meson-like operators, also meson-meson constructions has already been explored in two-point functions \cite{Dudek:2012gj, Dudek:2012xn, Dudek:2014qha}, and the corresponding three-point function calculations can now be attempted.


\begin{acknowledgements}
We thank our colleagues within the Hadron Spectrum Collaboration, in particular we acknowledge the assistance of R.A.~Brice\~no, D.J.~Wilson and C.E.~Thomas. CJS thanks B.J.~Owen for useful comments.  {\tt Chroma}~\cite{Edwards:2004sx} and {\tt QUDA}~\cite{Clark:2009wm,Babich:2010mu} were used to perform this work on clusters at Jefferson Laboratory under the USQCD Initiative and the LQCD ARRA project. We acknowledge resources used at Oak Ridge Leadership Computing Facility, the National Center for Supercomputing Applications, the Texas Advanced Computer Center and the Pittsburgh Supercomputer Center. Support is provided by U.S. Department of Energy contract DE-AC05-06OR23177 under which Jefferson Science Associates manages Jefferson Lab, the Early Career award contract DE- SC0006765, and the JSA Graduate Fellowship program.   
\end{acknowledgements}

\appendix

\section{Rotations and helicity operators\label{app::helOps}}

In this analysis we have made use of sets of kinematically equivalent helicity matrix elements. The operators appearing at the source and sink of our three point functions are subduced, with the subduction coefficients implying certain choices of rotation conventions. One method to consistently relate kinematically equivalent subduced matrix elements to a canonical frame involves embedding additional phases into the correlation functions appearing in our linear system, Equation \ref{eqn::linear_system}. Following \cite{Thomas:2011rh} a \emph{helicity state} is defined by 
\begin{equation}
\big|   \vec{p}; J,\lambda   \big\rangle \equiv \hat{R}_{\hat{p}}\,  \hat{L}_z(p) \big|  J,\lambda  \big\rangle 
\label{hel_defn}
\end{equation}
where $\big|J,\lambda \big\rangle$ is a state at rest with spin $J$ and $J_z$ component $\lambda$, $\hat{L}_z(p)$ is a boost along the $z$-axis with momentum magnitude $p$, and $\hat{R}_{\hat{p}}$ is a rotation that rotates from the $z$-axis to direction $\hat{p}$. A set of conventions to implement these rotations, which separates them into a rotation from the $z$-axis to a reference momentum direction, followed by a rotation which leaves the cubic lattice invariant is presented in \cite{Thomas:2011rh}, where the corresponding subduction into irreducible representations of the little group is also shown. 

For an arbitrary rotation, $R$, these helicity states transform as 
\begin{equation}
\hat{R}\, \big| \vec{p}; J,\lambda \big\rangle = e^{i \Phi(R, \vec{p}, J, \lambda)}\, \big| R \vec{p}; J,\lambda \big\rangle,
\label{hel_phase}
\end{equation}
where the helicity is left invariant and where rotations about the direction of the momentum, $\vec{p}$, introduce a helicity-dependent phase. 

A \emph{canonical state}, where the spin state is specified using the projection along the $z$-axis is defined as 
\begin{align*}
\big| \vec{p}; J,m \big\rangle &\equiv \hat{R}_{\hat{p}}\,  \hat{L}_z(p)\,  \hat{R}_{\hat{p}}^{-1}\, \big| J,m \big\rangle \\
&=  \hat{L}(\vec{p}) \, \big| J,m \big\rangle, 
\end{align*}
where $ \hat{L}(\vec{p})$ is a Lorentz boost along direction $\hat{p}$ with momentum magnitude, $p$. These states transform under rotations as 
\begin{equation*}
\hat{R} \,\big| \vec{p}; J,m \big\rangle = \sum_{m'} D^{(J)}_{m'm}\big(R\big) \, \big| R\vec{p}; J,m' \big\rangle,
\end{equation*}
and are related to helicity states via
\begin{equation}
\big| \vec{p}; J,\lambda \big\rangle  = \sum_{m} D^{(J)}_{m\lambda}\big(R_{\hat{p}}\big) \, \big| \vec{p}; J,m \big\rangle, \label{eqn::app_hel_to_can}
\end{equation}
where $R_{\hat{p}}$ is the rotation for direction $\hat{p}$ appearing in Eq.~\ref{hel_defn}.

The application of this formalism that is required for the work reported on in this paper is to consistently relate helicity matrix elements in different frames. To achieve this we must determine the phases in Eq.~\ref{hel_phase}, $e^{i \Phi(R, \vec{p}, J, \lambda)}$ corresponding to the rotation conventions laid down in \cite{Thomas:2011rh} used in the construction of our helicity operators. This can be achieved by evaluating matrix elements of the form $\big\langle \vec{p}\,'; J', \lambda' \big| \, \hat{R} \, \big|\,\vec{p}; J,\lambda \big\rangle $. It is convenient to exchange the helicity states for canonical states using Equation \ref{eqn::app_hel_to_can} -- the matrix element can then be written as 
\begin{align*}
\big\langle \vec{p}\,'&; J', \lambda' \big| \, \hat{R} \, \big|\,\vec{p}; J,\lambda \big\rangle  \\ 
&= \sum_{m',m}\! D^{(J')*}_{m' \lambda'}\big(R_{\hat{p}'} \big)\,  \big\langle \vec{p}\,'; J', m' \big| \, \hat{R} \, \big| \,\vec{p}; J,m \big\rangle    \, D^{(J)}_{m\lambda}\big(R_{\hat{p}} \big)
\\
&=\sum_{m',m}\! D^{(J')*}_{m' \lambda'}\big(R_{\hat{p}'} \big) \,  D^{(J)}_{m' m}(R) \,  D^{(J)}_{m\lambda} \big(R_{\hat{p}}\big) \; \delta^3\big(\vec{p}\,' - R\,\vec{p}\, \big) \;\delta_{J'J}.
\end{align*}

Using the composition of rotations we have
\begin{equation*}
D^{(J)}_{\lambda'\lambda}\big(R^{-1}_{\hat{p}'} R R_{\hat{p}} \big) = \sum_{m',m} D^{(J)*}_{m' \lambda'}\big(R_{\hat{p}'}\big)\,   D^{(J)}_{m' m}\big(R\big)  \, D^{(J)}_{m\lambda}\big(R_{\hat{p}} \big), 
\end{equation*}
and we conclude that the phase in Eq.~\ref{hel_phase} is given by
\begin{equation}
e^{i \Phi(R, \vec{p}, J, \lambda)} = D^{(J)}_{\lambda\lambda}\big(R^{-1}_{R\hat{p}} R  R_{\hat{p}} \big). \label{hel_phase_D}
\end{equation}

A generic matrix-element of the vector current between states of definite momentum, spin and helicity, is
\begin{equation*}
\big\langle \vec{p}\,'; J' ,\lambda'  \big| \, j^\mu\,  \big|\, \vec{p}; J, \lambda \big\rangle,
\end{equation*}
and for an arbitrary rotation $R$ this matrix element transforms as 
\begin{align*}
\big\langle \vec{p}\,'; J'& ,\lambda'  \big| \, j^\mu \, \big| \, \vec{p}; J, \lambda  \big\rangle  \\
 &= \big\langle \vec{p}\,'; J' ,\lambda' \big| \, \hat{R}^{-1}\hat{R} \, j^\mu \,  \hat{R}^{-1}\hat{R} \, \big| \, \vec{p}; J, \lambda  \big\rangle \\
&= \left[R^{-1}\right]^\mu_\nu    \big\langle \vec{p}\,'; J' ,\lambda'  \big|\,  \hat{R}^{-1} \, j^\nu\,  \hat{R} \, \big| \, \vec{p}; J, \lambda  \big\rangle,
\end{align*}
since the current rotates as a four-vector, ${\hat{R} \, j^\mu \, \hat{R}^{-1} = \big[ R^{-1} \big]^\mu_\nu \, j^\nu}$. Inserting complete sets of states, and using the fact, expressed in Eq.~\ref{hel_phase}, that rotations do not change the helicity of states, we obtain
\begin{widetext}
	\begin{equation}
		\big\langle \vec{p}\,'; J' ,\lambda'  \big| \, j^\mu\,  \big| \, \vec{p}; J, \lambda  \big\rangle = 
		\big[ R^{-1} \big]^\mu_\nu\; \int\!\!  d^3 q \int\!\!  d^3 k  \; 		
		\big\langle \vec{p}\,'; J',\lambda' \big| \, \hat{R}^{-1} \, \big|\, \vec{q}; J', \lambda' \big\rangle	 
		\big\langle \vec{q}; J', \lambda'   \big| \, j^\nu \,        \big|\, \vec{k}; J, \lambda   \big\rangle 
		\big\langle \vec{k}; J, \lambda     \big| \, \hat{R}\,       \big|\, \vec{p}; J, \lambda   \big\rangle.
	\end{equation}
Using our newly derived representation of the phase, Eq.~\ref{hel_phase_D}, we can write this as
\begin{equation*}
\big\langle \vec{p}\,'; J', \lambda'  \big| \, j^\mu \, \big| \, \vec{p}; J,\lambda \big\rangle = \left[R^{-1}\right]^\mu_\nu\, 
D^{(J')*}_{\lambda' \lambda'}\big( R^{-1}_{R\hat{p}'} R R_{\hat{p}'} \big)\;  
D^{(J)}_{\lambda \lambda}    \big( R^{-1}_{R\hat{p}} R R_{\hat{p}}   \big) \big\langle  R\vec{p}\,'; J', \lambda' \big|\, j^\nu \, \big|\,  R\vec{p}; J, \lambda \big\rangle 
\end{equation*}

In practice we use this expression to relate matrix elements determined for various allowed lattice momenta, $\vec{p}, \, \vec{p}\,'$, to matrix elements for some reference momenta $\vec{p}_\mathrm{ref} = R\vec{p},\,  \vec{p}\,'_{\!\mathrm{ref}} = R \vec{p}\,'$.

\pagebreak
\section{Momentum conservation in a finite-volume\label{app::two-point}}

\end{widetext}

We define meson eigenstates which in infinite volume have normalization
\begin{equation}
 \big\langle \mathfrak{n}(\vec{k}) \big| \mathfrak{n}'(\vec{p}) \big\rangle = \delta_{\mathfrak{n}\mathfrak{n}'}(2\pi)^3 \, 2 E_{\mathfrak{n}}(\vec{k}) \; \delta^{(3)}( \vec{k} - \vec{p} \,), \nonumber
\end{equation}
such that the completeness relation takes the form
\begin{equation}
1 = \sum_\mathfrak{n} \int \!\! \frac{d^3 \vec{k} }{(2\pi)^3} \, \frac{1}{2  E_{\mathfrak{n}}(\vec{k})  } 
	\big| \mathfrak{n}(\vec{k}) \big\rangle \big\langle \mathfrak{n}(\vec{k}) \big|.\nonumber
\end{equation}
In a periodic cubic volume, $L \times L \times L$, the allowed momenta of free particles is quantized, $\vec{k} = \tfrac{2\pi}{L} \vec{n}_k$, where $\vec{n}_k = \big(n_x, n_y, n_z \big)$ and the normalization  becomes $\big\langle \mathfrak{n}(\vec{k}) \big| \mathfrak{n}'(\vec{p}) \big\rangle = \delta_{\mathfrak{n}\mathfrak{n}'} \, 2 E_{\mathfrak{n}}(\vec{k}) \, L^3 \; \delta_{n_{\vec{k}},n_{\vec{p}}  }$, and completeness is expressed as
\begin{equation}
1 = \frac{1}{L^3} \sum_\mathfrak{n} \sum_{\vec{n}_k} \, \frac{1}{2 E_{\mathfrak{n}}(\vec{k}) } 
	\big| \mathfrak{n}(\vec{k}) \big\rangle \big\langle \mathfrak{n}(\vec{k}) \big|. \label{completeness}
\end{equation}

Two-point correlation functions in which the source and sink operators are projected into definite momentum have a spectral representation which can be obtained by inserting Eq.~\ref{completeness},
\begin{widetext}
\begin{align*}
C(t) 	&= \big\langle 0 \big| \mathcal{O}^{\,}_\mathrm{f}(\vec{p}_{\mathrm{f}}, t) \, \mathcal{O}^\dag_\mathrm{i}(\vec{p}_\mathrm{i}, 0) \big| 0 \big\rangle \\ 
		&= \big\langle 0 \big| \sum\nolimits_{\vec{x}} e^{i \vec{p}_\mathrm{f}\cdot \vec{x} }\,  \mathcal{O}^{\,}_\mathrm{f}(\vec{x}, t)\, 
		\sum\nolimits_{\vec{y}} e^{-i \vec{p}_\mathrm{i}\cdot \vec{y} }\,  \mathcal{O}^{\dag}_\mathrm{i}(\vec{y}, 0) \big| 0 \big\rangle \\
		&= \frac{1}{L^3} \sum_\mathfrak{n} \sum_{\vec{n}_k} \, \frac{1}{2 E_\mathfrak{n}(\vec{k}) } 
		\,\big( L^3 \delta_{\vec{p}_\mathrm{f} \vec{k}} \,\big)
		\,\big( L^3 \delta_{\vec{p}_\mathrm{i} \vec{k}} \,\big) 
		e^{-E_{\mathfrak{n}} t} \; 
		\big\langle 0 \big| \mathcal{O}_\mathrm{f}(\vec{x}=\vec{0}, 0) \big| \mathfrak{n}(\vec{k}) \big\rangle
		\big\langle \mathfrak{n}(\vec{k}) \big| \mathcal{O}^\dag_\mathrm{i}(\vec{y}=\vec{0}, 0) \big| 0\big\rangle \\
		&= L^3 \delta_{\vec{p}_\mathrm{f} , \vec{p}_\mathfrak{i}} \sum_\mathfrak{n} \frac{1}{2 E_\mathfrak{n} } 
		e^{-E_{\mathfrak{n}} t} \; 
		\big\langle 0 \big| \mathcal{O}_\mathrm{f}(\vec{0}, 0) \big| \mathfrak{n}(\vec{p}_\mathrm{i}) \big\rangle
		\big\langle \mathfrak{n}(\vec{p}_\mathrm{i}) \big| \mathcal{O}^\dag_\mathrm{i}(\vec{0}, 0) \big| 0\big\rangle,
\end{align*}
where we note an explicit factor of the lattice volume, $L^3$. 

Three-point correlation functions projected into definite source, sink and current momentum have a spectra representation,
\begin{align*}
C(t) 	&= 	\big\langle 0 \big| \mathcal{O}^{\,}_\mathrm{f}(\vec{p}_\mathrm{f}, t_\mathrm{f}) \,
			j(\vec{q}, t) \, 
			\mathcal{O}^\dag_\mathrm{i}(\vec{p}_\mathrm{i}, t_\mathrm{i}) \big| 0 \big\rangle \\
		&= \big\langle 0 \big| \sum\nolimits_{\vec{x}} e^{i \vec{p}_\mathrm{f}\cdot \vec{x} }\, \mathcal{O}^{\,}_\mathrm{f}(\vec{x}, t_\mathrm{f})\, 
		\sum\nolimits_{\vec{z}} e^{-i \vec{q} \cdot \vec{z} }\, j(\vec{z}, t)\,
		\sum\nolimits_{\vec{y}} e^{-i \vec{p}_\mathrm{i}\cdot \vec{y} }\, \mathcal{O}^{\dag}_\mathrm{i}(\vec{y}, t_\mathrm{i}) \big| 0 \big\rangle \\ 
		&= L^3 \delta_{\vec{p}_\mathrm{f} , \vec{p}_\mathfrak{i}+\vec{q}} \sum_{\mathfrak{n}_\mathrm{i},\mathfrak{n}_\mathrm{f} } \frac{1}{2 E_{\mathfrak{n}_\mathrm{i}} }\frac{1}{2 E_{\mathfrak{n}_\mathrm{f}} } 
		e^{-E_{\mathfrak{n}_\mathrm{f}} (t_f - t)} e^{-E_{\mathfrak{n}_\mathrm{i}} (t - t_i)} \; 
		\big\langle 0 \big| \mathcal{O}_\mathrm{f}(\vec{0}, 0) \big| \mathfrak{n}_\mathrm{f}(\vec{p}_\mathrm{f}) \big\rangle
		\big\langle \mathfrak{n}_\mathrm{f}(\vec{p}_\mathrm{f}) \big| j(\vec{0},0) \big| \mathfrak{n}_\mathrm{i}(\vec{p}_\mathrm{i}) \big\rangle
		\big\langle \mathfrak{n}_\mathrm{i}(\vec{p}_\mathrm{i}) \big| \mathcal{O}^\dag_\mathrm{i}(\vec{0}, 0) \big| 0\big\rangle,	
\end{align*}
which again features an explicit factor of the lattice volume, $L^3$. This volume factor, common to two-point and three-point functions may conventionally be absorbed into the meson creation/annihilation matrix elements.
\end{widetext}

\section{Vector Current Improvement on Anisotropic Clover Lattices,\label{app::Improv}}
\newcommand{\nablar}[1]{\overrightarrow{D}_#1}
\newcommand{\nablal}[1]{\overleftarrow{D}_#1}

The form of the anisotropic Clover fermion action that we use \cite{Lin:2008pr, Chen:2000ej} can be obtained from the desired Euclidean action $\bar{\Psi} ( m + \slashed{D} ) \Psi$ using the field transformations
\begin{align*}
	\Psi &= \big( 1 + \tfrac{1}{2} \Omega_m \, a_t m + \tfrac{1}{2} \Omega_t \, a_t \gamma_4 \overrightarrow{D}_{\!4} + \tfrac{1}{2} \Omega_s \, a_s \gamma_j \overrightarrow{D}_{\!j} \big)\psi \\
	\bar{\Psi} &= \bar{\psi} \big( 1 + \tfrac{1}{2} \bar{\Omega}_m \, a_t m + \tfrac{1}{2} \bar{\Omega}_t \, a_t \gamma_4 \overleftarrow{D}_{\!4} + \tfrac{1}{2} \bar{\Omega}_s \, a_s \gamma_j \overleftarrow{D}_{\!j} \big),
\end{align*}
leading to
\begin{align*}
	\bar{\psi} \Big[\;
		&\big( 1 + \tfrac{1}{2}(\Omega_m + \bar{\Omega}_m) a_t m \big) m \\
		&\;+ \big( 1 + \tfrac{1}{2}(\Omega_m + \bar{\Omega}_m) a_t m + \Omega_t \, a_t m \big) \gamma_4 \overrightarrow{D}_{\!4}  + \Omega_t \, a_t \overrightarrow{D}_{\!4}\overrightarrow{D}_{\!4} \\
		&\;+ \big( 1 + \tfrac{1}{2}(\Omega_m + \bar{\Omega}_m) a_t m + \Omega_s \, a_s m \big) \gamma_j \overrightarrow{D}_{\!j}  + \Omega_s \, a_s \overrightarrow{D}_{\!j}\overrightarrow{D}_{\!j} \\
		&\quad+ \tfrac{1}{2}\big( \Omega_t a_t + \Omega_s a_s \big) \sigma_{i4} F_{i4} + \big(\Omega_s a_s \big) \sum_{i>j} \sigma_{ij} F_{ij} \;\;
	 \Big]\psi
\end{align*}
where a term proportional to $\big\{ D_i, D_4 \big\}$ was eliminated by choosing $\bar{\Omega}_t = - \Omega_t$ and $\bar{\Omega}_s = - \Omega_s$. Making the choices, $\bar{\Omega}_m + \Omega_m = 1$, $\Omega_s = -\tfrac{1}{2} \nu_s$, $\Omega_t = -\tfrac{1}{2}$, discretizing the derivatives, smearing in the spatial directions and inserting tadpole factors gives the action presented in \cite{Edwards:2008ja, Lin:2008pr}. Reference \cite{Edwards:2008ja} describes the non-perturbative tuning of the parameters.

The same transformation can be applied to the desired vector current $j_\mu = \bar{\Psi} \gamma_\mu \Psi$ to obtain the classically $O(a)$ improved vector current. The transformation gives
\begin{align*}
	j_\mu &= \big(1 + \tfrac{1}{2} a_t m \big)\, \bar{\psi} \gamma_\mu \psi    \\
	&\quad- \tfrac{1}{4} a_t \big( \partial_4 (\bar{\psi} \sigma_{\mu 4} \psi)  - \delta_{\mu 4} \bar{\psi} ( \overleftarrow{D}_{\!4} - \overrightarrow{D}_{\!4})\psi \big) \\
	&\quad- \tfrac{1}{4} \nu_s a_s \big( \partial_j (\bar{\psi} \sigma_{\mu j} \psi)  - \delta_{\mu j} \bar{\psi} ( \overleftarrow{D}_{\!j} - \overrightarrow{D}_{\!j})\psi \big),
\end{align*} 
and use of the classical equations of motion for the quark fields allows for the elimination of the gauge-covariant derivatives acting on quark fields to give
\begin{align*}
j_4 &= \big( 1 + \tfrac{1}{2}(m+m_0)a_t \big)\,  \bar{\psi}\gamma_4 \psi + \tfrac{1}{4} \tfrac{\nu_s}{\xi} ( 1- \xi) \, a_s \partial_j \big( \bar{\psi}\sigma_{4j} \psi\big)  \\
j_k &= \big( 1 + \tfrac{1}{2}(m+m_0 \xi)a_t \big)\,  \bar{\psi}\gamma_k \psi + \tfrac{1}{4} ( 1 - \xi) \, a_t \partial_4 \big( \bar{\psi}\sigma_{4k} \psi\big),
\end{align*}
where $\xi = a_s/a_t$ is the anisotropy.

Since we non-perturbatively determine the renormalization of the vector current at any given quark mass by computing meson form-factors at zero momentum transfer, we choose to absorb the mass-dependent factor into the renormalization factor and consider the improved vector currents,
\begin{align}
j_4 &= Z_V^t \Big(   \bar{\psi}\gamma_4 \psi + \tfrac{1}{4} \tfrac{\nu_s}{\xi} ( 1- \xi) \, a_s \partial_j \big( \bar{\psi}\sigma_{4j} \psi\big)  \Big) \nonumber \\
j_k &= Z_V^s \Big(   \bar{\psi}\gamma_k \psi + \tfrac{1}{4} ( 1 - \xi) \, a_t \partial_4 \big( \bar{\psi}\sigma_{4k} \psi\big) \Big).
\end{align}
As expected we observe that the improvement terms vanish at classical level in the case of an isotropic action, $\xi = 1$.
 

\section{Wick contractions for three quark flavors\label{app::flavour}}

\begin{figure}[t]
\includegraphics[width=\linewidth]{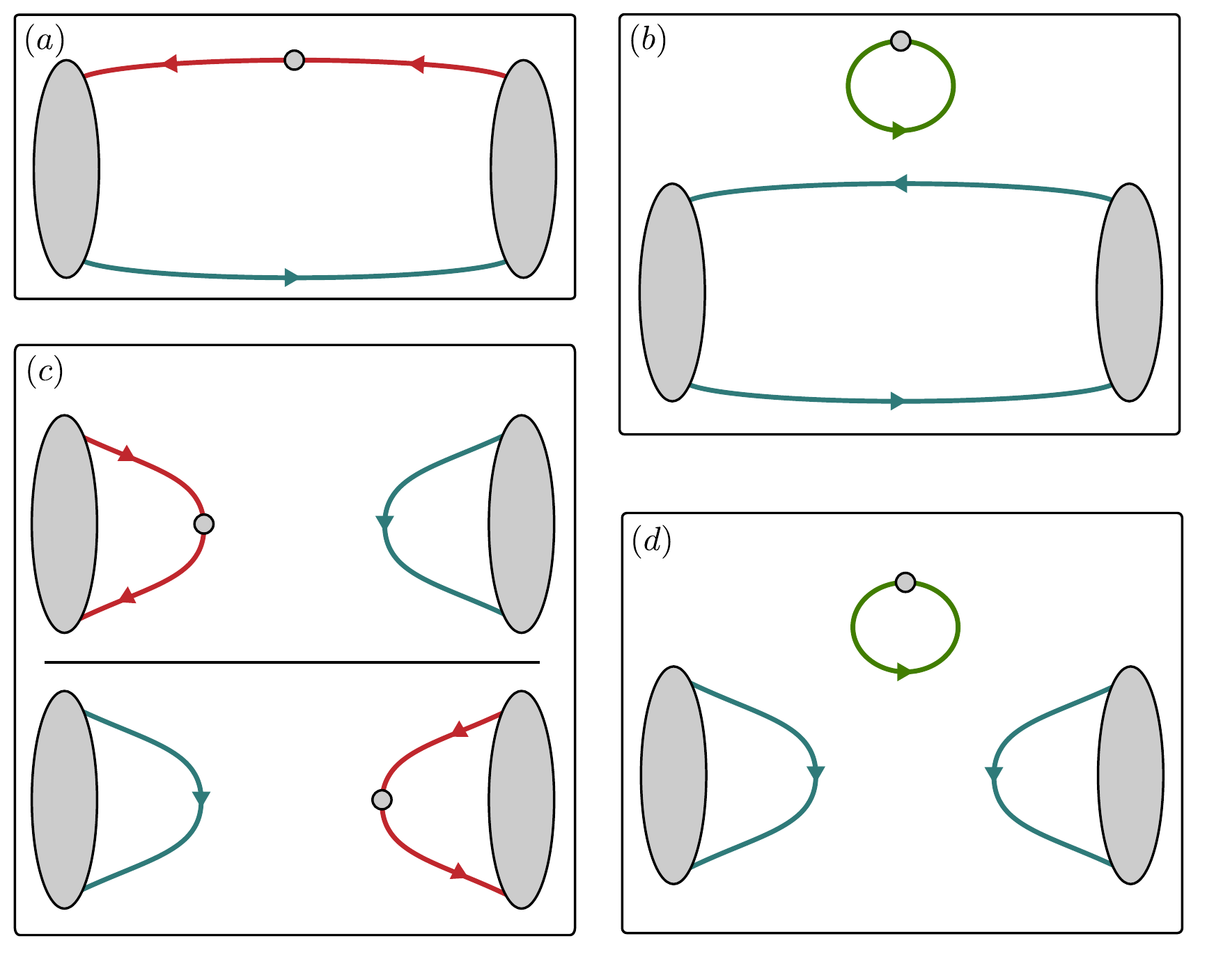}
\caption{Possible quark line contractions required for meson three-point functions with a current insertion. Blue lines represent quark propagation which could be described by a perambulator and red lines represent quark propagation with a current insertion which could be described by a generalized perambulator. Green lines indicate quark propagation which does not begin or end on an operator which can be distillation smeared, and as such cannot be described by a perambulator. \label{wicks} }
\end{figure}

In this paper we work with a version of QCD in which the up, down and strange quarks all have the same mass, leading to an exact $SU(3)_F$ symmetry. We compute radiative transition matrix elements between states in octets, $\mathbf{8}_F$, of $SU(3)_F$. In particular we consider the $(I,I_z) = (1,\pm 1)$ elements, and interpolate mesons from the vacuum with operators having flavor structure $\bar{d} \mathbf{\Gamma} u$.

Formally integrating out fermions from the path-integral we replace field pairs $q_x\bar{q}_y$ with the appropriate quark propagator, $Q_{xy}$, the inverse of the Dirac operator. Since the Dirac operator depends upon the quark mass, there can be differing propagators for each flavor of quark, i.e. $u\bar{u} \to U$, $d\bar{d} \to D$, $s\bar{s} \to S$. In the case of exact isospin symmetry, $m_u = m_d$ and $U=D$ and if $SU(3)_F$ is exact, $U=D=S$.

For transitions between $I=1$ mesons, inserting currents $\bar{u} \gamma_\mu u$, $\bar{d} \gamma_\mu d$ or $\bar{s} \gamma_\mu s$, the schematic Wick contractions shown in Figures~\ref{wicks}(a) and (b) appear, where those shown in Figure~\ref{wicks}(b) are referred to as \emph{disconnected} contributions. Were we to consider $I=0$ mesons, we could also receive contributions from the Wick contractions shown in Figures~\ref{wicks}(c) and (d).

The electromagnetic current operator in units of $e$ has the flavor structure ${j^\mathrm{em}_\mu = q_u \bar{u} \gamma_\mu u + q_d \bar{d} \gamma_\mu d + q_s \bar{s} \gamma_\mu s}$, and it is convenient to express this in terms of $SU(3)_F$ singlet and octet currents,
\begin{align}
j^\mathrm{em}_\mu = \quad \tfrac{1}{\sqrt{3}} (q_u + q_d + q_s)\, &j^{\mathbf{1}-}_\mu \nonumber \\
					+\tfrac{1}{\sqrt{6}} (q_u + q_d - 2q_s)\, &j^{\mathbf{8}-}_\mu \nonumber \\
					+\tfrac{1}{\sqrt{2}} (q_u - q_d)\, &j^{\mathbf{8}+}_\mu,
\end{align}
where
\begin{align}
	j^{\mathbf{1}-}_\mu &= \tfrac{1}{\sqrt{3}}\left( \bar{u} \gamma_\mu u + \bar{d} \gamma_\mu d   + \bar{s} \gamma_\mu s\right) \nonumber\\
	j^{\mathbf{8}-}_\mu &= \tfrac{1}{\sqrt{6}}\left( \bar{u} \gamma_\mu u + \bar{d} \gamma_\mu d  -2 \bar{s} \gamma_\mu s\right) \nonumber \\
	j^{\mathbf{8}+}_\mu &= \tfrac{1}{\sqrt{2}}\left( \bar{u} \gamma_\mu u - \bar{d} \gamma_\mu d  \right), \label{SU3current} 
\end{align}
and where the $\pm$ superscript indicates the current's \mbox{$G$-parity}.

Considering Eq.~\ref{SU3current} in the $SU(3)_F$ limit where the propagators for up, down and strange quarks are identical ($U=D=S\equiv Q$), it is clear that the octet currents $j^{\mathbf{8}+}_\mu$ and $j^{\mathbf{8}-}_\mu$ cannot give rise to a disconnected contribution of the type shown in Figure~\ref{wicks}(b), as the relevant part of the Wick contraction would be proportional to 
\begin{align*}
\mathrm{tr}\left[\gamma_\mu U \right] - \mathrm{tr}\left[\gamma_\mu D \right] &=\mathrm{tr}\left[\gamma_\mu Q \right] - \mathrm{tr}\left[\gamma_\mu Q \right] = 0 \nonumber \\
\mathrm{tr}\left[\gamma_\mu U \right] + \mathrm{tr}\left[\gamma_\mu D \right] - 2\, \mathrm{tr}\left[\gamma_\mu S \right] &=(1+1-2)\mathrm{tr}\left[\gamma_\mu Q \right] = 0,
\end{align*}
respectively. In general, the singlet current could give rise to a disconnected contribution proportional to $\mathrm{tr}\left[\gamma_\mu Q \right]$, but since it enters the electromagnetic current with a weight of $q_u + q_d + q_s = +\tfrac{2}{3} - \tfrac{1}{3} - \tfrac{1}{3} = 0$, it too does not contribute in practice.

For the case of the form-factors of $(I,I_z) = (1,+1)$ mesons like the $\pi^+$ or $\rho^+$, the initial and final state mesons are identical and have the same $G$-parity. The \mbox{$G$-parity} invariance of QCD (assuming isospin invariance) then ensures that the $G$-parity of the current must be positive and only $j^{\mathbf{8}+}_\mu$ can contribute. As such the matrix element is proportional to $q_u - q_d = +\tfrac{2}{3} - -\tfrac{1}{3} = +1$ as expected, and we see that there cannot be any disconnected diagrams even away from the $SU(3)_F$ limit\footnote{although there can be if isospin symmetry is broken by $m_u \neq m_d$}.

For transitions like $\rho^+ \to \pi^+ \gamma$, where the $G$-parity flips, while there are disconnected contributions away from the $SU(3)_F$ limit, the logic presented above indicates that there are not in this calculation which is performed with exact $SU(3)F$ symmetry. Only connected diagrams are required, and the matrix element is proportional to $q_u + q_d$. We note that in the $SU(3)_F$ case, the matrix element for $K^{\star+} \to K^+ \gamma$ is identical to $\rho^+ \to \pi^+ \gamma$.

\bibliographystyle{apsrev4-1}
\bibliography{bib}

\end{document}